\documentclass[a4paper,aps,prb,twocolumn,superscriptaddress]{revtex4-2}

\usepackage{amsmath,amssymb,graphicx,bbold}
\usepackage{braket}
\usepackage{xcolor}
\usepackage[colorlinks=true, urlcolor=blue, linkcolor=blue, citecolor=blue, pdftex]{hyperref}
\usepackage{orcidlink}
\usepackage{multirow}

\graphicspath{ {./figures/} {./sketches/} }

\def\<{\langle}
\def\>{\rangle}

\newcommand{\ve}[1]{\boldsymbol{#1}}
\newcommand{\id}{\mathbb{1}}

\newcommand{\cdag}[1]{\hat{c}_{#1}^\dag}
\newcommand{\cd}[1]{\hat{c}_{#1}^{\phantom{\dag}}}
\newcommand{\nop}[1]{\hat{n}_{#1}}

\newcommand{\JMP}{J.~Mat.~Phys.}
\newcommand{\NPB}{Nucl.~Phys.~B}
\newcommand{\PRX}{Phys. Rev. X}

\newcommand{\suna}{\ensuremath{\mathfrak{su}(N)}}
\newcommand{\sun}{\ensuremath{\text{SU}(N)}}

\begin{document}
\title{Phase diagram of the SU($N$) antiferromagnet of spin $S$ on a square lattice}
\author{\firstname{Jonas} \surname{Schwab}\,\orcidlink{0000-0003-3794-8631}}
\email{jonas.schwab@physik.uni-wuerzburg.de}
\affiliation{Institut f\"ur Theoretische Physik und Astrophysik and W\"urzburg-Dresden Cluster of Excellence ct.qmat, Universit\"at W\"urzburg, 97074 W\"urzburg, Germany}
\author{\firstname{Francesco} \surname{Parisen Toldin}\,\orcidlink{0000-0002-1884-9067}}
\email{parisentoldin@physik.rwth-aachen.de}
\affiliation{Institute for Theoretical Solid State Physics, RWTH Aachen University, Otto-Blumenthal-Str. 26, 52074 Aachen, Germany}
\affiliation{JARA-FIT and JARA-CSD, 52056 Aachen, Germany}
\author{\firstname{Fakher F.} \surname{Assaad}\,\orcidlink{0000-0002-3302-9243}}
\email{assaad@physik.uni-wuerzburg.de}
\affiliation{Institut f\"ur Theoretische Physik und Astrophysik and W\"urzburg-Dresden Cluster of Excellence ct.qmat, Universit\"at W\"urzburg, 97074 W\"urzburg, Germany}

\begin{abstract}
We investigate the ground state phase diagram of  an   \sun-symmetric antiferromagnetic  spin model on a square lattice  where each site 
hosts an  irreducible representation of \sun\ described by a square Young tableau  of $N/2$ rows and $2S$ columns.  We  show  that    negative sign 
free  fermion  Monte  Carlo simulations can be  carried  out for this class of   quantum magnets  at  any  $S$ and  even values of $N$. 
 In the large-$N$ limit,  the  saddle point  approximation 
favors  a   four-fold   degenerate    valence  bond  solid phase.  In the large $S$-limit,   the  semi-classical  approximation points  to   N\'eel  state.   
On a  line  set  by    $N=8S +  2 $  in the $S$ versus  $N$  phase  diagram,  we  observe  a  variety of phases proximate to the   N\'eel state.   At  $S = 1/2 $ and  $3/2 $   we  observe the   aforementioned   four fold  degenerate   valence  bond solid  state.  At  $S=1$  a  two  fold  degenerate  spin  nematic  state in which  the  C$_4$ lattice  symmetry  is  broken down to  C$_2$ emerges. Finally  at  $S=2$  we  observe a  unique  ground  state  that pertains  to a  two-dimensional  version of  the  Affleck-Kennedy-Lieb-Tasaki  state.    For  our  specific  realization, this   symmetry  protected  topological  state   is  characterized  by  an SU(18), $S=1/2$  boundary  state,  that  has  a  dimerized  ground state.       These  phases that are  proximate  to the 
N\'eel  state  are  consistent  with the notion of  monopole  condensation of  the  antiferromagnetic  order parameter.  In particular    one  expects    spin disordered  states  with degeneracy  set  by   $\text{mod}(4,2S)$. 
 \end{abstract}

\maketitle

\section{Introduction}
\label{sec:intro}
Spin systems are ubiquitous in nature and form one of the most fundamental concept in condensed matter and statistical physics.
Their complex collective behavior has spurred numerous experimental and theoretical studies, aimed at understanding their nature and properties.
At the same time, modeling of spin systems represents a primary theoretical laboratory to investigate fundamental physics.
Starting   with the  classical Ising model \cite{Onsager44}, spin systems   have  played  a crucial role  in our  understanding  of 
phase  transitions \cite{Sachdev_book},   phases  of  matter, frustration and disorder \cite{Diep_book},   emergent  gauge  theories \cite{Castelnovo12,Balents10}  
and    exotic    critical    behavior   \cite{Senthil04_2}.
The impact of spin models extends beyond the realm of condensed matter physics, and has found application in other areas, such as information processing \cite{Nishimori-book} and quantum computing \cite{Nielsen-Chuang_book}, where the fundamental unit of information, a qubit, is a single spin-$1/2$ system.

 \begin{figure}[!h]
  \centering
  \includegraphics[width=0.95\linewidth]{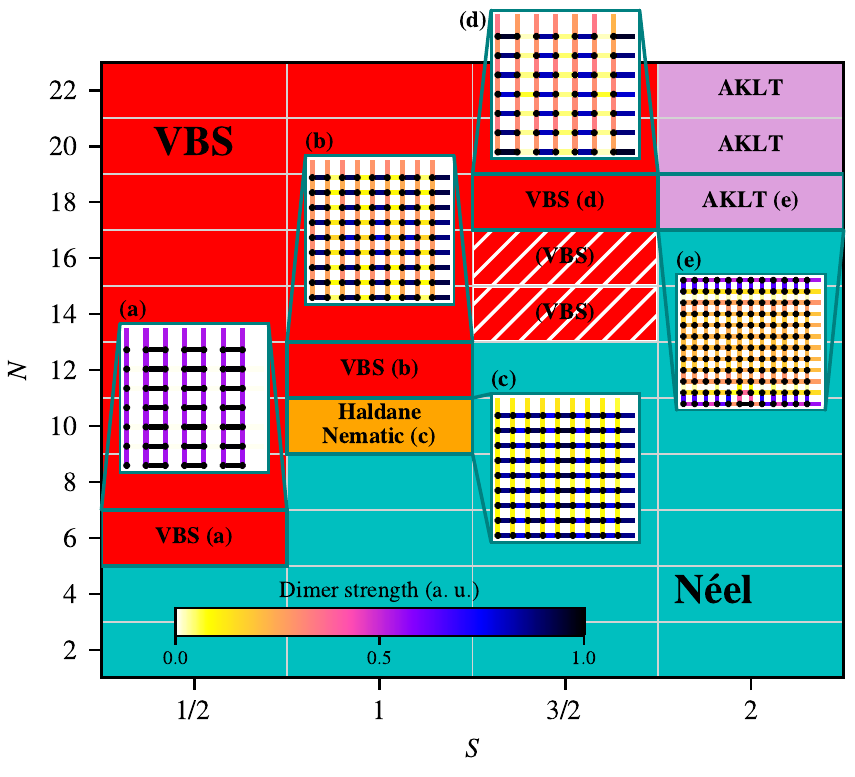}
  \caption{
    Ground-state phase diagram of the \sun-antiferromagnet model (\ref{HAF}) on the square lattice, as obtained from QMC simulations. $S$ identifies the chosen representation of the \suna\ algebra of \sun, illustrated by the Young tableau in Fig.~\ref{young}.
    Striped regions indicate the part of the phase diagram where current QMC data do not allow an unambiguous identification of the phase;
    in such cases we indicate between parenthesis the most likely identified order.
    The insets show QMC data in the highlighted dimerized phases, obtained through a pinning-field approach (see Sec.~\ref{sec:results:op}).
  }
  \label{fig:PhaseDiagram}
 \end{figure}

In condensed matter, 
spin systems   are realized in  Mott insulators, which arise when  charge fluctuations in a given unit cell are suppressed.  
For instance, in undoped cuprates the copper atom is in a Cu$^{2+}$ state and corresponds to a net spin $S=1/2$ degree of freedom. Super-exchange leads to an $S=1/2$, SU(2) Heisenberg spin model that has been studied numerically \cite{Sandvik97,Calandra98} and experimentally \cite{Coldea01} at length. Higher spin SU(2) systems arise when $2S$ electrons are localized on a single orbital and a strong Hund's rule favors a maximal spin state with a totally symmetric wave function. 
For example, in the Haldane chain realized by the CsNiCl$_3$ compound Ni$^{2+}$ ions carry spin 1 \cite{Zaliznyak50}.  
\sun-invariant models, for $N>2$, naturally arise  as special cases of the Kugel-Khomski model \cite{Kugel82,Kugel15}, where spin and orbital degrees of freedom turn out to play a very symmetric role.  
In particular, the observed spin-orbital liquid behavior in Ba$_3$CuSb$_2$O$_9$ \cite{Nakatsuji12} has been interpreted in terms of an SU(4) quantum antiferromagnet in the defining representation \cite{Corboz12}. 
Beyond the solid state physics, \sun\ spin models can be realized in the realm of cold atomic gases \cite{Wu03,Gorshkov10}. 

\begin{figure}
  \centering
  \includegraphics[width=0.7\linewidth]{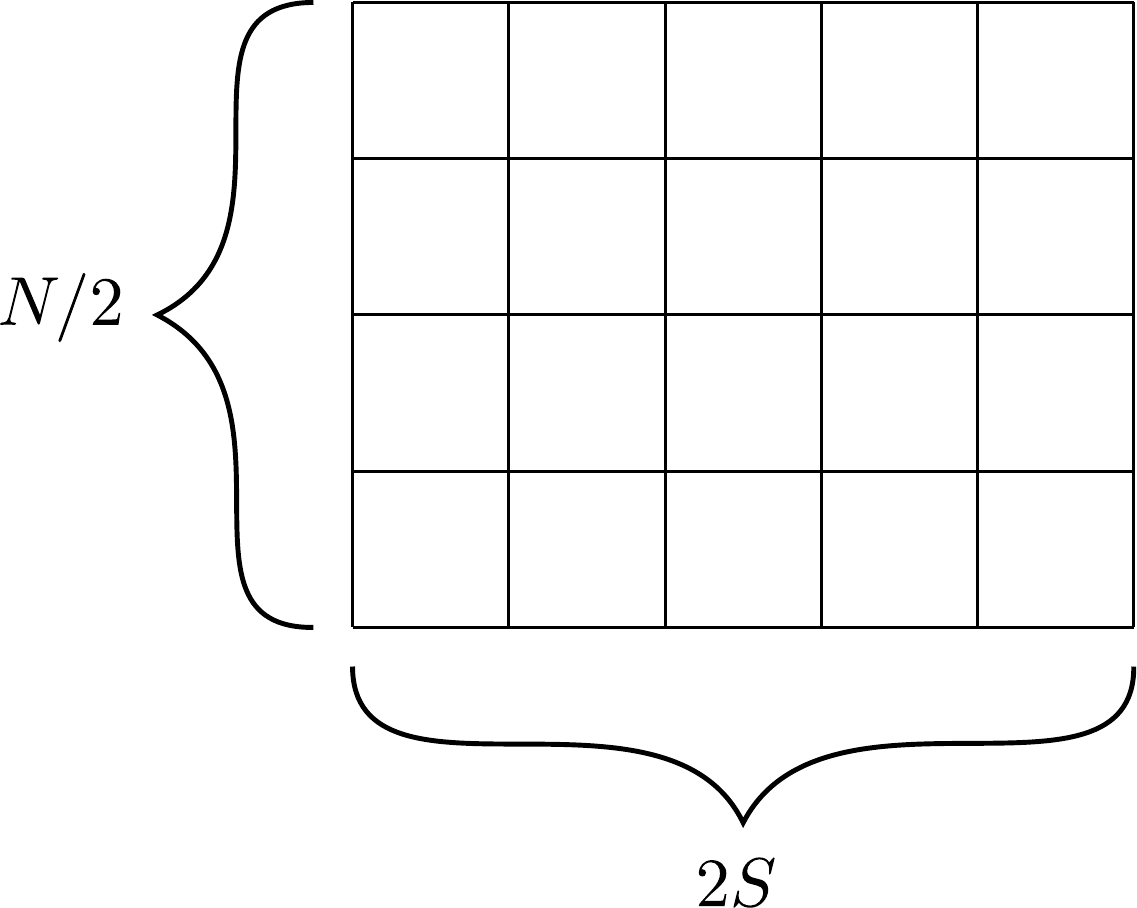}
  \caption{Young tableau corresponding to the irreducible representation of \suna\ considered here; $N$ is even and $S$ is semi-integer.}
  \label{young}
\end{figure}

Topology plays a decisive role in the understanding of SU(2) invariant spin systems. In fact, using a spin coherent-state  path integral approach to 
antiferromagnetic (AFM)  Heisenberg chains,  one  
identifies  a  Berry  phase.   It  corresponds to the  skyrmion   count  of  the  three-components  normalized  order  parameter
in 1+1 dimensions  and  at   angle   $\theta=  2\pi S$ \cite{Haldane83}.  This  provides  a  topological understanding    of  the  observed  
differences  between   half-integer   and  integer  spin  chains.     In two   spatial  dimensions,   topology    enters    through singular    skyrmion  number 
  changing  events  in  space  time:  monopoles   \cite{Haldane88}.     For  a  square  lattice  with $C_4$ symmetry,  only quadrupole  (double) monopole 
  events  are  allowed  for    half-integer  (odd)   spin by  symmetry.   There is no constraint  on the monopole number  for  even values of  the spin.   For  the   plain
  vanilla  SU(2)   Heisenberg model  at   arbitrary   spin $S$, the  spin-wave  approximation captures  well  the  ground  state  and   topological excitations lie 
  high in the spectrum.   In this  context,  the  theory  of   deconfined  quantum  criticality  essentially  poses  the  question of  the nature of the quantum 
  phase transition that emanates  when    one  decreases  the  energy  of  monopoles  and  ultimately  condenses  them \cite{Senthil04_2,Senthil04_2}.  
    For  half-integer  spin  systems,   where 
  only  quadrupole  monopole  insertions are allowed,  one  can conjecture  that  the  Hilbert  space   splits  into   four orthogonal  subspaces  characterized 
  by  the  number of monopoles modulo  four.    This  provides  an understanding of  how the    fourfold   degenerate  valence  bond  solid (VBS)     state  emerges    
  for  condensing topological  excitations  of  the  quantum antiferromagnet  \cite{Sandvik07}.   Similarly,   for spin-1 (spin-2)  systems,   condensing  monopoles  should    generate a  twofold (zerofold) degenerate   disordered state.   
  	
		  A  crucial  question is  how  to    control the  monopole  energy.   The seminal work of  Read and Sachdev \cite{RS-89,RS-89b,RS-90}  
 shows  that the  discussion above    can be  carried over to \sun\ spin systems, $N\ge 2$. Furthermore,  enhancing  $N$  has  the potential  of  lowering the monopole  energy.     In this  paper,  
 we  show  that it is  possible to formulate  negative sign-free  auxiliary field (AF) quantum Monte Carlo (QMC)  simulations  \cite{BSS-81,WSSLGS-89,Sorella89,AF_notes,ALF_v2} of the \sun\ AFM  spin-$S$  Heisenberg 
 model,   for  representations    given by a Young    tableau with $N/ 2$   rows  and  $2S$  columns.    
 This  generalizes the work of Ref.~\cite{Assaad04} to generic values of $S$. Specifically, we  consider  the model: 
 \begin{equation}
  \hat{H} = J\sum_{\langle i, j  \rangle, a } \hat{S}^{(a)}_i\hat{S}^{(a)}_j,
  \label{HAF}
\end{equation}
where the sum extends over the pair of nearest-neighbor sites $\langle i, j \rangle$,   and  $a$ runs over   $N^2 - 1$  generators  of
the  said  representation of the \suna\ algebra of \sun. 
The main result of this paper is the rich phase diagram illustrated in Fig.~\ref{fig:PhaseDiagram}.
Remarkably,  and for  each considered  value  of  $S=1/2,1,3/2,2$    just above  the   threshold   value of  $N$ above  which N\' eel order  disappears,  we  observe   four-, two- and zerofold degenerate    disordered  states    at   half-integer, odd  and even values of  $S$.

This paper is organized as follows.
In Sec.~\ref{sec:formulation} we discuss how we construct the Hamiltonian (\ref{HAF}), with the spin operators in the desired representation of Fig.~\ref{young}.
In Sec.~\ref{sec:qmc} we illustrate its actual implementation within a fermionic representation, which can be sampled by means of QMC simulations in the AF approach.
In Sec.~\ref{sec:results} we present and discuss our QMC results for the phase diagram of the model.
In Sec.~\ref{sec:summary} we summarize our findings. 
In Appendix~\ref{app:casimir_young} we discuss a formula giving the eigenvalue of the quadratic Casimir operator of a representation in terms of its Young tableau.
In Appendix~\ref{app:casimir_max} we prove an upper bound on the eigenvalue of the Casimir operator of the irreducible representations emerging from a tensor product of representations discussed in Sec.~\ref{sec:formulation}.
In Appendix~\ref{app:dtau_errors} we discuss the systematic error in the QMC formulation, arising from the Trotter discretization.
In Appendix~\ref{app:bounds} we prove a lower and upper bound for a bond observable used to diagnose the phases.

\section{General  formulation of the  Hamiltonian}  

\label{sec:formulation}
In the Hamiltonian (\ref{HAF}),
the operators $S^{(a)}_i$   form  an irreducible representation of the \suna\ algebra.
This is uniquely specified by its maximum Dynkin weight $\Lambda_{\alpha}$ or, alternatively, by a Young tableau, from which the components $\Lambda_{\alpha_k}$ can be read off as \cite{Vergados_book}
\begin{equation}
  \Lambda_{\alpha_k} = l_k - l_{k+1},\qquad k=1,\ldots, N-1,
  \label{young_to_dynkin}
\end{equation}
where $l_k$ is the length of the $k$-th row of the Young tableau, and one can assume $l_N=0$ for representations of \suna.

Here we consider, on each lattice site, the representation corresponding to a Young tableau illustrated in Fig.~\ref{young}, whose corresponding maximum Dynkin weight is
\begin{equation}
  \Lambda_{\alpha_k}=2S\delta_{k,N/2}.
  \label{dynkin}
\end{equation}
The dimension of an irreducible representation can be computed with the hook-length formula, or with Weyl's formula \cite{Vergados_book}:
\begin{equation}
  \text{dim} = \prod_{i<j}^N \frac{l_i-l_j+j-i}{j-i}.
  \label{weyl_dimension}
\end{equation}
For the present case, we have
\begin{equation}
  \text{dim}=\prod_{j=0}^{N/2-1}\frac{\left(2S+\frac{N}{2}+j\right)! j!}{\left(2S+j\right)!\left(\frac{N}{2}+j\right)!}.
  \label{dimension_rep}
\end{equation}

To realize this representation,
we first introduce on each lattice site
$2S$ independent irreducible representations.
Their tensor product decomposes into different irreducible representations, including, in particular, the one of Fig.~\ref{young}.
In a second step, we project the Hilbert space onto that of the desired representation by maximizing the quadratic Casimir operator.

Let $\{T_a\}$, $a=1,\ldots, N^2-1$ be a basis of the \suna\ algebra.
We start by introducing on each lattice site $i$ the antisymmetric self-adjoint representation $T_a\rightarrow \Gamma(T_a) = \hat{T}_{a,i}$.
Its maximum weight in the Dynkin representation is
\begin{equation}
  \Lambda_{\alpha_k} = \delta_{k,N/2},
\label{dynkin_S1_2}
\end{equation}
which matches Eq.~(\ref{dynkin}) for $S=1/2$.
Equivalently, in agreement with Eq.~(\ref{young_to_dynkin}), this representation corresponds to a Young tableau with one column and $N/2$ boxes.

\begin{figure}
  \centering
  \includegraphics[width=\linewidth]{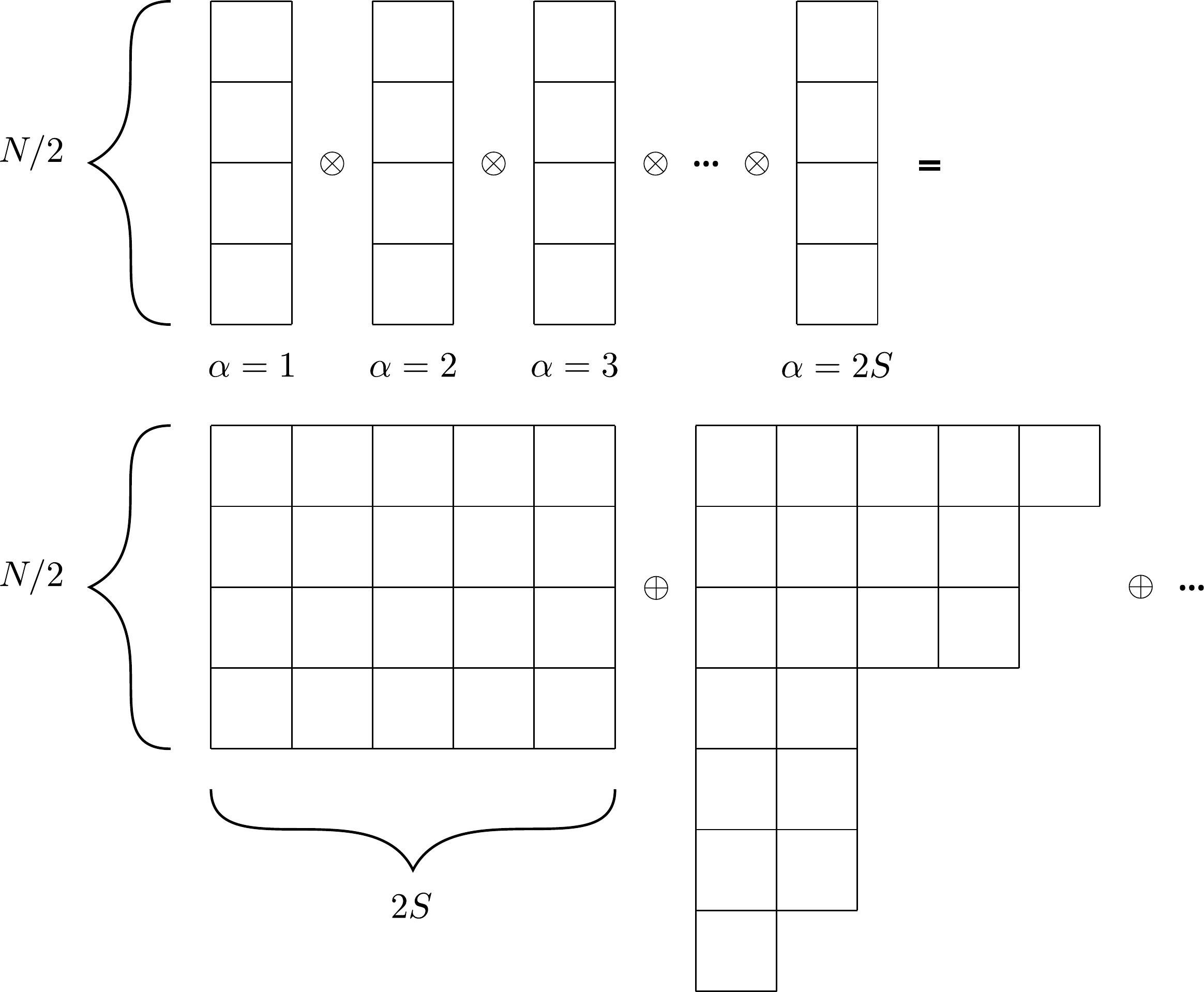}
  \caption{Decomposition of the tensor product of $2S$ antisymmetric self-adjoint representations, whose maximum Dynkin weight is given in Eq.~(\ref{dynkin_S1_2}), into irreducible ones.}
  \label{decomposition}
\end{figure}

Next, we consider, for each lattice site $i$, $2S$ independent representations $\hat{T}_{a,i,\alpha}$, $\alpha=1\ldots 2S$.   We  refer 
to  $\alpha$ as the flavor index.
The composite generators, i.e., the generators for the tensor product of the $2S$ representations, define the spin operators appearing in Eq.~(\ref{HAF}) and are given by
\begin{equation}
  \hat{S}_i^{(a)} = \sum_{\alpha=1}^{2S} \hat{T}_{a,i,\alpha},
  \label{spin_rep}
\end{equation}
Using Eq.~(\ref{spin_rep}), the interaction term in Eq.~(\ref{HAF}) is written as \footnote{In Eq.~(\ref{HAF}) and Eq.~(\ref{HQMC_int}) we have implicitly assumed a choice of the basis of \suna, such that the interaction term is \sun-invariant. This condition is satisfied by the basis choice given below in Eq.~(\ref{Ta_norm}).}
\begin{equation}
\begin{split}
  \hat{H}_J &= J \sum_{\langle i\ j  \rangle} \sum_{a=1}^{N^2-1}\hat{S}^{(a)}_i \hat{S}^{(a)}_j \\
  &= J \sum_{\langle i\ j  \rangle} \sum_{a=1}^{N^2-1}\sum_{\alpha,\beta=1}^{2S} \hat{T}_{a,i,\alpha}\hat{T}_{a,j,\beta}.
  \label{HQMC_int}
\end{split}
\end{equation}
The operators $\hat{S}_i^{(a)}$ in Eq.~(\ref{spin_rep}) form a reducible representation of \suna, which
decomposes into several irreducible representations, illustrated in Fig.~\ref{decomposition}.
As proven in App.~\ref{app:casimir_max}, among the resulting representations, the one of Fig.~\ref{young} exhibits the maximum eigenvalue of the quadratic Casimir operator.
To explicitly compute it, we choose a basis $\{T_a\}$ of \suna\ such that
\begin{equation}
  \text{Tr}\{T_aT_b\} = \frac{1}{2}\delta_{ab}.
  \label{Ta_norm}
\end{equation}
With this choice, the structure constants of the algebra are completely antisymmetric and the chosen basis is, up to a trivial normalization, self-dual with respect to the bilinear form (\ref{Ta_norm}). Thus, given an irreducible representation $\Gamma: \suna\rightarrow GL(d,\mathbb{C})$, we define the quadratic Casimir operator as
\begin{equation}
  \hat{C}_2 = \sum_a \Gamma(T_a)\Gamma(T_a) \equiv C\id_d,
  \label{casimir_def}
\end{equation}
where we have used the fact that $C_2\propto \id_d$ (Schur's Lemma) to introduce the eigenvalue of the Casimir operator $C$
\footnote{We notice that the operator defined in Eq.~(\ref{casimir_def}) commutes with the algebra only for completely antisymmetric structure constants. For a general choice of the base, one needs to introduce a metric tensor $g_{ab}$ determined by the structure constants and the Casimir operator is defined as $\sum_{ab}g^{ab}\Gamma(T_a)\Gamma(T_b)$ \cite{Vergados_book}; for completely antisymmetric structure constants $g_{ab}\propto\delta_{ab}$. Also, for the same reason, a normalization is implicit in the definition of Eq.~(\ref{casimir_def}), discussed in App.~\ref{app:casimir_young}.}.
Using Eq.~(\ref{spin_rep}) in Eq.~(\ref{casimir_def}), the quadratic Casimir operator on the lattice site $i$ is
\begin{equation}
  \hat{C}_{2,\Gamma_i} = \sum_{a=1}^{N^2-1}\sum_{\alpha,\beta=1}^{2S}\hat{T}_{a,i,\alpha}\hat{T}_{a,i,\beta}
    \label{casimir_total}
\end{equation}

In order to project the Hilbert space to the subspace of the desired representation, we introduce on each site
a term in the Hamiltonian which favors the states with the highest Casimir value
\begin{equation}
  \begin{split}
    \hat{H}_{\text{Casimir}} &= -J_H\sum_i \hat{C}_{2,\Gamma_i}\\
    &= -J_H \sum_i\sum_{a=1}^{N^2-1}\sum_{\alpha,\beta=1}^{2S}\hat{T}_{a,i,\alpha}\hat{T}_{a,i,\beta},
  \end{split}
  \label{HQMC_casimir}
\end{equation}
with $J_H>0$.
The term of Eq.~(\ref{HQMC_casimir}) effectively introduces a ferromagnetic interaction between different flavors, with coupling strength $J_H$.

The  Hamiltonian  that  we    will  solve  numerically,  reads: 
\begin{equation}
    \hat{H} =   \hat{H}_J  + \hat{H}_{\text{Casimir}}. 
\end{equation}
Importantly,   $  \left[  \hat{H}_J, \hat{H}_{\text{Casimir}} \right] = 0 $,  such that the projection  
onto the desired  irreducible  representation  turns out to be  very  efficient.   
And since the projection is a local onsite term, we expect it to scale independent from system size.

\section{QMC  formulation}
\label{sec:qmc}
\subsection{Fermionic representation}
\label{sec:qmc:fermionic}
As discussed in Sec.~\ref{sec:formulation}, the Hamiltonian is constructed using as basic building blocks antisymmetric self-adjoint representations, defined by the maximum weight of Eq.~(\ref{dynkin_S1_2}) or, equivalently, by a Young tableau with one column and $N/2$ boxes.
The corresponding operators $\hat{T}_{a,i,\alpha}$ entering in Eqs.~(\ref{HQMC_int}) and (\ref{HQMC_casimir})
can be realized by introducing, for every lattice site $i$ and for every flavor index $\alpha$,
$N$ nonrelativistic fermions, with creation and annihilation operators $\cdag{i,\alpha,\sigma}$, $\cd{i,\alpha,\sigma}$, $\sigma=1\ldots N$, and fixing the total charge (i.e., the number of fermions) to half-filling, i.e., to $N/2$.
For every $i$ and $\alpha$,
a basis of this Hilbert space is generated by the states
\begin{equation}
    \left( \cdag{i,\alpha,1}\right)^{n_{i,\alpha,1}} \cdots   \left(\cdag{i,\alpha,N}\right)^{n_{i,\alpha,N}}  | 0 \rangle,\qquad n_{i,\alpha,\sigma} = 0,1,
\end{equation}
with the constraint
\begin{equation}
  \sum_{\sigma=1}^{N}  n_{i,\alpha,\sigma} = \frac{N}{2}, \qquad \forall i, \alpha.
  \label{charge_constraint}
\end{equation}
In this space, the 
(representation of the) \suna\ generators are
\begin{equation}
  \hat{T}_{a,i,\alpha} = \sum_{\sigma,\sigma'}\cdag{i,\alpha,\sigma}(T_a)_{\sigma\sigma'}\cd{i,\alpha,\sigma'}.
  \label{antisym_rep}
\end{equation}
It is easy to check that the maximum weight in the Dynkin representation agrees with Eq.~(\ref{dynkin_S1_2}), thus providing us the needed building block to simulate the Hamiltonian (\ref{HAF}).

We study the model by means of  finite-temperature AF QMC  \cite{AF_notes,BSS-81,WSSLGS-89}  and  projective AF QMC
\cite{Sugiyama86,Sorella89,AF_notes}.  In this framework, we  sample  respectively  the  grand canonical  and canonical  ensembles at  half-filling, and charge fluctuations are generally present.
 Therefore, we need to additionally impose the constraint of Eq.~(\ref{charge_constraint}).
Notice that, unlike available techniques for canonical QMC simulations \cite{WAPT-17,SLYR-20}, where the global charge of the system is fixed, here we need to impose half-filling on each lattice site.
To this end, we add  a repulsive Hubbard $U$-term on each site $i$ and flavor $\alpha$:
\begin{equation}
  \hat{H}_U = U\sum_i \sum_{\alpha=1}^{2S}\left(\hat{n}_{i,\alpha}-\frac{N}{2}\right)^2, \quad \nop{i,\alpha} \equiv \sum_{\sigma=1}^{N} \cdag{i,\alpha,\sigma}\cd{i,\alpha,\sigma}.
  \label{HQMC_Hub}
\end{equation}

\begin{figure}
  \centering
  \includegraphics[width=0.8\linewidth]{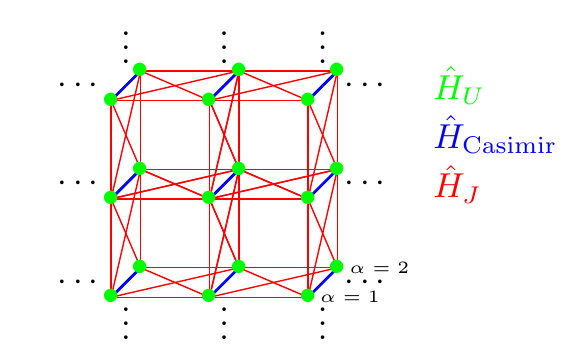}
  \caption{
    \label{fig:ham_sketch}
    Sketch of the structure of the QMC Hamiltonian for $2S=2$. {\color{green} $\hat{H}_U$}: Hubbard term for freezing out charge degrees of freedom. {\color{blue} $\hat{H}_\text{Casimir}$}: Term for maximizing the eigenvalue of the Casimir operator. {\color{red} $\hat{H}_J$}: Antiferromagnetic interaction between elemental spins.
  }
\end{figure}
In summary, the Hamiltonian simulated with the AF QMC method is the sum of the interaction term given in Eq.~(\ref{HQMC_int}), the Casimir term [Eq.~(\ref{HQMC_casimir})], and the Hubbard term [Eq.~(\ref{HQMC_Hub})], with the operators $\{\hat{T}_{a,i,\alpha}\}$ given in Eq.~(\ref{antisym_rep}).
Eqs.~(\ref{HQMC_int}) and (\ref{HQMC_casimir}) can be further simplified using the following summation identity \cite{Haber-21}
\begin{equation}
  \sum_{a=1}^{N^2-1} (T_a)_{\sigma\sigma'}(T_a)_{\epsilon\epsilon'} = \frac{1}{2}\left(\delta_{\sigma\epsilon'}\delta_{\sigma'\epsilon}-\frac{1}{N}\delta_{\sigma\sigma'}\delta_{\epsilon\epsilon'}\right),
  \label{sum_generators}
\end{equation}
which holds for a choice of generators that satisfies Eq.~(\ref{Ta_norm}). Using Eq.~(\ref{sum_generators}) and collecting the terms in Eqs.~(\ref{HQMC_int}), (\ref{HQMC_casimir}) and (\ref{HQMC_Hub}), the QMC Hamiltonian is
\begin{equation}
  \begin{split}
    \hat{H}_{\text{QMC}} =&\hat{H}_J + \hat{H}_{\text{Casimir}} + \hat{H}_U \\
    =&- \frac{J}{4}  \sum_{\langle i,j \rangle, \alpha,\beta } \left\{ \hat{D}_{(i,\alpha),(j,\beta)}, \hat{D}^{\dagger}_{(i,\alpha),(j,\beta)}   \right\} \\
    &+ \frac{J_H}{2} \sum_{i} \sum_{\alpha>\beta}    \left\{ \hat{D}_{(i,\alpha),(i,\beta)}, \hat{D}^{\dagger}_{(i,\alpha),(i,\beta)}   \right\}\\
    &+ U\sum_{i,\alpha}\left(\nop{i,\alpha}-\frac{N}{2}\right)^2,\\
  \end{split}
  \label{HQMC}
\end{equation}
where
\begin{equation}
  \hat{D}_{(i,\alpha),(j,\beta)} \equiv  \sum_{\sigma} \cdag{i,\alpha,\sigma}\cd{j,\beta,\sigma},
  \label{HQMC_D}
\end{equation}
$\left\{\hat{A}, \hat{B}\right\} \equiv \hat{A} \hat{B} + \hat{B} \hat{A}$, and $\nop{i,\alpha}$ as defined in Eq.~(\ref{HQMC_Hub}).
The Hamiltonian now takes the form of the Heisenberg model  considered in Ref.~\cite{Assaad04} and the   proof  for the  absence of sign problem  is similar.
In Fig.~\ref{fig:ham_sketch} we sketch the resulting interactions for the case $S=1$. 

Before  proceeding,  we  would like to comment on the computational cost  of  the AF QMC  algorithm  \cite{AF_notes} for  this model.    The   total  number  of orbitals  is   given   by  $L^2 2S $ such   that  matrix   operations   required  to  compute,  e.g.,   the  single-particle   spectral function,   scales  as 
$ (L^2 2S)^3 \beta  $,  where $\beta$ is  the inverse  temperature.     It  turns  out,  that,  in  contrast  to the generic Hubbard  model  with $L^2 2S$  sites,  this is not  the leading  computational cost.    The  number  of    Hubbard-Stratonovitch  fields per  imaginary  time  slice  scales as     $L^2  S^2$.  Using  fast   updates,   refreshing  one  field   involves    $(L^2 2S)^2$  floating point operations,  such  that the  total  cost of the  updating  scales  as  $L^6S^4\beta$.    Hence large  values of  $S$  are   computationally  expensive.   
In Appendix \ref{app:dtau_errors}, we  show  that the  computational cost  does not    explicitly scale with  $N$. 
We  note  that this estimate  of  the computational cost  does not take into account   auto-correlation  times.

\subsection{Test of projections}
\label{sec:qmc:projections}
As discussed above, the Hamiltonian (\ref{HQMC}) is equivalent to Eq.~(\ref{HAF}) in the limit $J_H\rightarrow\infty$, and $U\rightarrow\infty$, under which the Hilbert space is projected to the representation of Fig.~\ref{young}.
To optimally test the projections, we use the finite-temperature AF QMC method, which evaluates
  $\langle  \hat{O} \rangle  =   \text{Tr} \left[ e^{- \beta  \hat{H}}  \hat{O} \right]  /   \text{Tr} \left[ e^{- \beta  \hat{H}}  \right] $,   where   the   trace   runs over  the  grand canonical ensemble.

\begin{figure}
  \includegraphics[width=\linewidth]{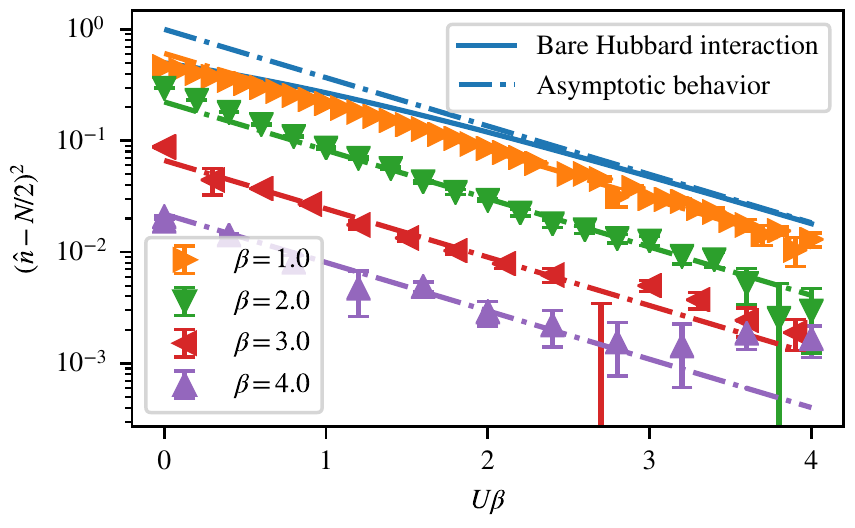}
  \caption{
    Suppression of the charge fluctuations $\<\left( \hat{n}_{i,1}- N/2 \right)^2\>$ as a function of $\beta U$, for $N=2$ and $S=1/2$.
    We consider a Hamiltonian containing the Hubbard term only and the case of  a model with an antiferromagnetic interaction [Eq.~(\ref{HQMC_int})], with coupling constant $J=1$ on a system size $L=4$, and for different inverse temperatures $\beta$.
    The charge fluctuations fall off asymptotically as $\exp(-\beta U )$ [Eq.~(\ref{charge_suppression})].
}
  \label{fig:Utest_N2}
\end{figure}

The interaction term of Eq.~(\ref{HQMC_int}) and the Casimir term of Eq.~(\ref{HQMC_casimir}) manifestly conserve the charge on each lattice site $i$.
Hence, in the Gibbs density matrix $\exp(-\beta\hat{H})$, the Hubbard term factorizes out, resulting in an effective exponential suppression of the charge fluctuations,
\begin{equation}
  \Braket{\left( \hat{n}_{i,\alpha}- N/2 \right)^2} \propto e^{- \beta U},
  \label{charge_suppression}
\end{equation}
independent from system size.
The suppression of charge fluctuations is therefore particularly efficient.
This is illustrated
in Fig.~\ref{fig:Utest_N2}, where we show $\<\left( \hat{n}_{i,1}- N/2 \right)^2\>$ in a semilogarithmic scale for $N=2$ and $S=1/2$ and as a function of $\beta U$.
Besides the case of a Hamiltonian containing the Hubbard interaction [Eq.~(\ref{HQMC_Hub})] only, for which any observable depends only on $\beta U$, we consider the presence of the AFM interaction Eq.~(\ref{HQMC_int})
for a lattice of linear size $L=4$. 
In the latter case, there is an additional dependence on the inverse temperature $\beta$, which we illustrate by considering four values.
In line with Eq.~(\ref{charge_suppression}), we observe an exponential suppression of the charge fluctuations as a function of $\beta U$.
Interestingly, in the interacting case, $\<\left( \hat{n}_{i,1}- N/2 \right)^2\>$ decreases with the temperature for any given value of $\beta U$, even for $U=0$.
This implies that the AFM coupling itself suppresses the charge fluctuations.

In Appendix~\ref{app:casimir_young}, we discuss a formula that gives the value of the Casimir eigenvalue in terms of the Young tableau of the representation \cite{PS-84}.
Employing this result, in Appendix~\ref{app:casimir_max} we determine, for the representation of Fig.~\ref{young}:
\begin{equation}
  C(N, S) = \frac{NS(2S+N)}{4}.
  \label{casimir_thisrep}
\end{equation}
Furthermore, in Appendix~\ref{app:casimir_max}, we prove that
Eq.~(\ref{casimir_thisrep}) is the maximum Casimir eigenvalue
among the irreducible representations arising from the tensor product of $2S$ self-adjoint antisymmetric representations given in Eq.~(\ref{antisym_rep}), and that
there is a finite gap $O(1)$ in the eigenvalues of the quadratic Casimir operators between the maximally symmetric representation of Fig.~\ref{young} and the other irreducible representations arising from the tensor product.
Therefore, the term of Eq.~(\ref{HQMC_casimir}) effectively selects a single representation, and the projection is efficient.

\begin{figure}
  \centering
  \includegraphics[width=\linewidth]{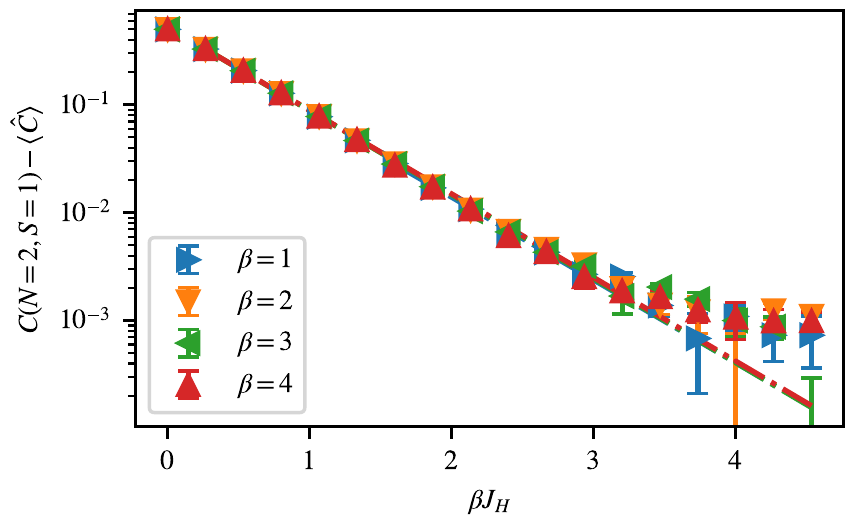}
  \caption{
    Difference between the Casimir eigenvalue $C(N=2, S=1)$ [Eq.~(\ref{casimir_thisrep})] of the representation $S=1$, $N=2$, and the sampled one $\<\hat{C}\>$, as a function of the effective interaction strength $\beta J_H$ [Eq.~(\ref{HQMC_casimir})] with $J=0$, and for four inverse temperatures.
    Data shown are obtained for a lattice of size $L=4$, with a Hubbard interaction $\beta U=6$ [Eq.~(\ref{HQMC_Hub})], vanishing nearest-neighbor antiferromagnetic interaction $J=0$, and a Trotter discretization $\Delta\tau=0.1$.
  }
  \label{fig:casimir_test_N2}
\end{figure}

\begin{figure}
  \includegraphics[width=\linewidth]{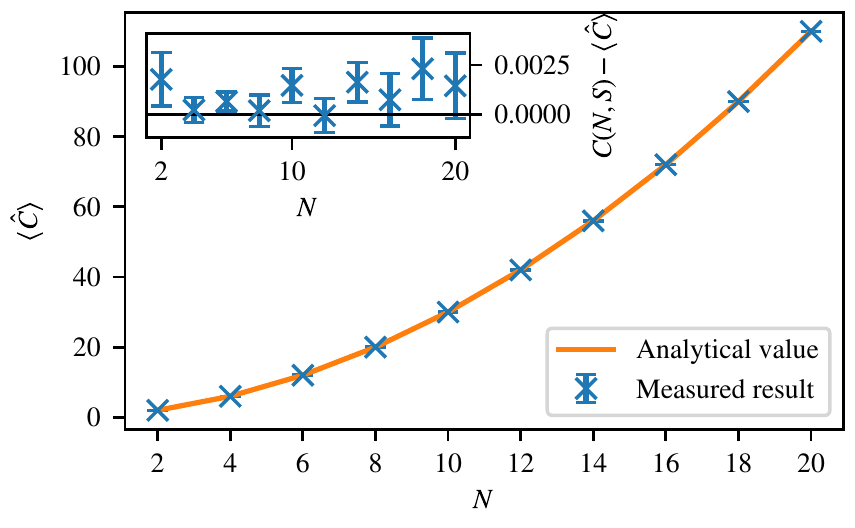}
  \caption{
    Casimir eigenvalue for representations with $S=1$ and as a function of $N$.
    We compare the predicted value of Eq.~(\ref{casimir_thisrep}) with the sampled Casimir eigenvalue from QMC simulations of a lattice with size $L=4$, with a Hubbard interaction $U=2$ [Eq.~(\ref{HQMC_Hub})], antiferromagnetic coupling $J=1$ [Eq.~(\ref{HAF})], projection strength $J_H=1$ [Eq.~(\ref{HQMC_casimir})], and a Trotter discretization $\Delta\tau=0.1$.
    In the inset we plot the difference between the sampled and expected value.
  }
  \label{fig:casimir_test_manyN}
\end{figure}

To control the projection, we compute the expectation value of the quadratic Casimir operator from the QMC simulations and compare it with the expected result of Eq.~(\ref{casimir_thisrep}).
An example of such a projection is shown in Figs.~\ref{fig:casimir_test_N2} and \ref{fig:casimir_test_manyN}.
In Fig.~\ref{fig:casimir_test_N2}, we plot the difference between the computed and expected Casimir eigenvalue $C$, as a function of $\beta J_H$, for different inverse temperatures and in a semilogarithmic scale.
The deviation from the expected result is exponentially suppressed in $\beta J_H$, underscoring the effectiveness of the projection.
In Fig.~\ref{fig:casimir_test_manyN}, we show, as a function of $N$, the sampled value of $C$ along with the expected result, and in the inset we plot their difference, which vanishes within error bars.

\section{Results}
\label{sec:results}
\subsection{Order parameters and phases}
\label{sec:results:op}
We have simulated the Hamiltonian Eq.~(\ref{HQMC}) using the ALF package \cite{ALF,ALF_v2},
which provides a comprehensive library to program QMC simulations of interacting models of fermions, using the AF algorithm \cite{AF_notes,BSS-81,WSSLGS-89}.
In particular, we used the projective formulation of the algorithm, which projects a trial wave function $\ket{\Psi_{\rm T}}$ onto the ground state of the system. Observables are evaluated through
\begin{align}
\label{eq:projetiveQMC2}
\braket{\hat{O}} &=
\frac{
  \Braket{\Psi_{\rm T}| e^{-\Theta\hat{H}}  
  \hat{O}
  e^{-\Theta\hat{H}} |\Psi_{\rm T}}
  }{   
  \Braket{\Psi_{\rm T}| e^{-2\Theta\hat{H}} |\Psi_{\rm T}}
  },
\end{align}
with $\Theta$ the projection parameter.
The algorithm employs a Hubbard-Stratonovich decomposition of the interaction terms.
This results in a free fermionic system, where any observable can be computed via the Wick's theorem from the Green's functions.
The QMC method consists in a stochastic sampling of the Hubbard-Stratonovich fields.
We refer to Ref.~\cite{AF_notes} for a discussion of the AF QMC method.

As trial wave function $\ket{\Psi_{\rm T}}$, we used the half-filled ground state of
\begin{align}
\hat{H}_{\rm T} = \sum_{\langle i,j \rangle} \sum_{\alpha=1}^{2S} 
  \left( \hat{D}_{(i,\alpha),(j,\alpha)} + {\rm H.c.} \right).
\end{align}
We scaled the projection parameter $\Theta$ with linear system size $L$, usually comparing the results obtained with $\Theta = L/4$ and $\Theta = L/2$, ensuring that they reflect ground state properties.
Furthermore, we chose the parameters for suppression of charge fluctuations and projection onto the maximally symmetric representation around $U= 4/\Theta$, $J_H = 4/\theta$, while always checking that charge fluctuations are sufficiently suppressed and $\braket{\hat{C}} = C(N, S)$ [cf. Eq.~\eqref{casimir_thisrep}].

To detect the realization of different ground states, we have sampled the spin two-point function $S(\ve{k})$ and the correlations of the dimer operator $D_{ij}(\ve{k})$ in momentum space, defined as
\begin{align}
S(\ve{k}) \equiv \frac{1}{(N^2 -1)N_r^2} &\sum_{\ve{r},a} e^{i \ve{k} \ve{r}}  \Braket{\hat{S}^{(a)}_{\ve{0}} \hat{S}^{(a)}_{\ve{r}}},
\label{Scorr}
\\
\begin{split}
D_{ij}(\ve{k}) \equiv \frac{1}{(N^2 -1)N_r^2} &\sum_{\ve{r},a,b} e^{i \ve{k} \ve{r}}\cdot
\\  
\Bigg[
&\Braket{\left(\hat{S}^{(a)}_{\ve{0}} \hat{S}^{(a)}_{\ve{0}+\ve{e}_i}\right)
          \left(\hat{S}^{(b)}_{\ve{r}} \hat{S}^{(b)}_{\ve{r}+\ve{e}_j}\right) }\\
  &- \Braket{\hat{S}^{(a)}_{\ve{0}} \hat{S}^{(a)}_{\ve{0}+\ve{e}_i}}
  \Braket{\hat{S}^{(b)}_{\ve{r}} \hat{S}^{(b)}_{\ve{r}+\ve{e}_j}} \Bigg],
\end{split}
\label{Dcorr}
\end{align}
where $N_r=L^2$ is the number of sites in a lattice of linear size $L$ and $\ve{e}_i$ is the elementary lattice unit vector on the $i-$th direction.
The normalization in Eqs.~(\ref{Scorr}) and (\ref{Dcorr}) ensure a finite thermodynamic and large-$N$ limit.
Using these observables, we can distinguish the N\'eel state and different dimerized ground states, to be discussed below.

The AFM N\'eel state exhibits long-range spin-spin correlations at momentum $\ve{k}=(\pi, \pi)$.
Thus, it can be detected by the staggered magnetization $m$
\begin{equation}
  m^2 = S(\ve{k}=(\pi,\pi)).
  \label{staggm}
\end{equation}

\begin{figure}
  \centering
  \includegraphics[scale=1]{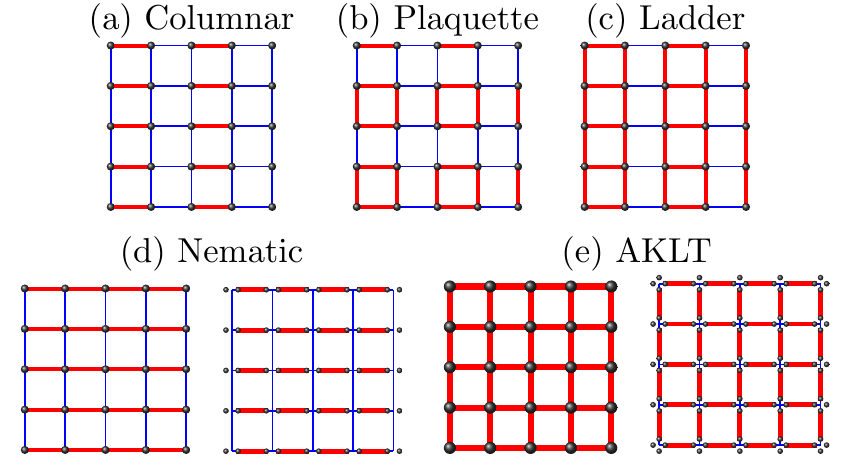}
  \caption{Sketch of possible dimerized ground states.
  (a)-(c) The VBS states. Each of those states is fourfold degenerate, the corresponding states can be obtained by rotations and translations.
  (d) A ``Haldane nematic'' state, an equivalent state is obtained by rotations of $90^{\circ}$.
  (e) A  unique ground  state.  
  States  (d) and (e)  are  at  best  understood  within  an AKLT construction,  in which,  very  much as  done in our  
  calculation, the  spin  $S$  on  each  site is constructed  by  a  totally   symmetric  superposition  of  $2S$ states   that are  denoted  by  bullets  
    around each  site on the  right-hand  side of  (d) and (e).  These  bullets  correspond  to an  irreducible   representation of  \suna\ with  one  column ($S=1/2$) and   $N/2$  rows.      In the nematic  state,   each  spin-$1/2$   forms  a singlet   with  the nearest  neighbor  along  the  axis of the  broken  symmetry.     The  AKLT  state  is   relevant  for  the  $S=2$  state,  where  each spin-$1/2$  on a  given  site  can be  combined into a  singlet  with  a nearest-neighbor  spin-$1/2$  without breaking a lattice  symmetry.
  }
  \label{fig:vbs_nematic}
\end{figure}
The valence bond state (VBS) breaks the lattice rotation and translation symmetries, realizing a fourfold degenerate pattern of strong and weak dimers.
This is realized by different sets of bond configurations, illustrated in Figs.~\ref{fig:vbs_nematic}(a)-\ref{fig:vbs_nematic}(c).
Beyond the commonly identified columnar order, sketched in Fig.~\ref{fig:vbs_nematic}(a), there are two additional VBS states [Figs.~\ref{fig:vbs_nematic}(b) and \ref{fig:vbs_nematic}(c)].
Notably, all three patterns break the lattice translation symmetry, but only columnar and ladder order break the four-fold rotation symmetry.
VBS order can be detected by a suitable order parameter $\phi$ defined in terms of the dimer correlations
\begin{equation}
  \phi^2 \equiv D_{xx}(\ve{k}) + D_{yy}(\ve{k}), \qquad \ve{k}=(\pi, 0).
  \label{phi}
\end{equation}
We average $\phi$ over the two equivalent momenta $(\pi,0)$, and $(0,\pi)$ as to obtain an improved estimator.

For integer values of $S$, we investigate the possible realization of the ``Haldane nematic'' Affleck-Kennedy-Lieb-Tasaki (AKLT) phase \cite{AKLT-88,RS-89,RS-89b,RS-90}.
This state is twofold degenerate and breaks the rotational symmetry but, unlike the VBS state, does not break translational symmetry.
We illustrate it in Fig.~\ref{fig:vbs_nematic}(d).
For such a phase we have $\phi=0$ and a suitable order parameter can be defined as \cite{DK-19}
\begin{equation}
  \psi^2 \equiv D_{xx}(\ve{k}) + D_{yy}(\ve{k}) - D_{xy}(\ve{k}) - D_{yx}(\ve{k}), \qquad \ve{k}=\ve{0}.
  \label{psi}
\end{equation}
$\psi$ is designed to pick up rotation symmetry breaking in the dimers and therefore does also not vanish for columnar and ladder order. 
Therefore, $\psi$   distinguishes  plaquette  with $C_4$ symmetry   from  VBS   order  with  broken   $C_4$  symmetry.

Finally, a two-dimensional version of the AKLT phase, with singlets on all bonds as sketched in Fig.~\ref{fig:vbs_nematic}(e), is also possible for $S=2$. This non-degenerate state does not break any  symmetry,  therefore all previously defined order parameters vanish.
In Table \ref{order_parameters} we summarize the different  orders. 
\begin{table*}
  \caption{List of considered ground states with their ordering momenta in reciprocal space and matrix of order parameters defined in Eqs.~(\ref{staggm}), (\ref{phi}), (\ref{psi}).}
  \begin{ruledtabular}
    \begin{tabular}{llccccc}
      Phase &
      Ordering momenta &
      $C_4$ Lattice symmetry preserved &
      $m$ &
      $\phi$ &
      $\psi$ \\
      \hline
      N\'eel    & $(\pi, \pi)$                               & yes &  $\ne 0$ & $0$     & $0$ \\
      Columnar  & $(\pi, 0)$ or  $(0, \pi)$  & no & $0$     & $\ne 0$ & $\ne 0$ \\
      Plaquette & $(\pi, 0)$ and $(0, \pi)$               & yes &   $0$     & $\ne 0$ & $0$ \\
      Ladder    &  $(\pi, 0)$ or  $(0, \pi)$  & no &  $0$     & $\ne 0$ & $\ne 0$ \\
      Nematic   & $(0, 0)$                                & no &  $0$     & $0$     & $\ne 0$ \\
      2d AKLT   & $(0, 0)$                                   & yes & $0$     & $0$     & $0$
    \end{tabular}
  \end{ruledtabular}
  \label{order_parameters}
\end{table*}

Given a local order parameter $O$  at momentum $\ve{p}$, one can define the correlation ratio $R_O$ as
\begin{equation}
  R_O \equiv 1 - \frac{C_O(\ve{p}+\delta\ve{p})}{C_O(\ve{p})},
  \label{corr_ratio}
\end{equation}
where $C_O(\ve{p})$ is the two-point function of the order parameter in Fourier space, and $\delta\ve{p}$ is the minimum nonzero momentum on a finite lattice.
On the square lattice, $\delta\ve{p}=(2\pi/L,0)$ or $\delta\ve{p}=(0,2\pi/L)$; as usual, one can average over the two minimum displacements to obtain an improved estimator.
The correlation ratio is closely related to the second-moment finite-size correlation length $\xi$, which on a square lattice can be defined as \cite{CP-98,PTHAH-14}
\begin{equation}
  \xi = \frac{1}{2\sin (\pi/L)}\sqrt{\frac{C_O(\ve{p})}{C_O(\ve{p}+\delta\ve{p})}-1}.
  \label{xi}
\end{equation}
  In a disordered phase, $\xi$ as defined in Eq.~(\ref{xi}) converges to the second-moment correlation length for $L\rightarrow\infty$, such that $\xi/L\rightarrow 0$ and $R_0\rightarrow 0$.
  In an ordered phase, due to the lack of spontaneous symmetry breaking in any finite size, $\xi/L$ diverges for $L\rightarrow \infty$ and, conversely, $R_0\rightarrow 1$.
In the vicinity of a critical point, $R_O$ and $\xi/L$ are renormalization-group invariant quantities.
Their crossing can be used to locate the onset of the phase transition, rendering them powerful quantities to diagnose the ground-state order and to study phase transitions.

An ergodic QMC simulation  averages over  all  symmetry-breaking states.  As a result, we are not able to observe the ordered state directly, but have to refer to correlation functions that do not average out to zero when averaging over all degenerate ground states. Unfortunately, such an approach does not distinguish between the different VBS states illustrated in Figs.~\ref{fig:vbs_nematic}(a)-\ref{fig:vbs_nematic}(c). To obtain additional insights we use the method of a pinning field \cite{AH-13,PTAW-17}. In this approach, we explicitly break the symmetry by making one AFM interaction at the origin $J_{\text{pin}}\sum_a \hat{S}^{(a)}_{\ve{0}} \hat{S}^{(a)}_{\ve{0}+\ve{e}_{\rm x}}$ stronger than the other interactions $J\sum_a \hat{S}^{(a)}_{\ve{r}} \hat{S}^{(a)}_{\ve{r}+\ve{e}_i}$.   The  resulting  Hamiltonian reads: 
\begin{align}
\hat{H} &= J \sum_{(\ve{r}, i) \neq (\ve{0}, {\rm x})} \sum_{a}
   \hat{S}^{(a)}_{\ve{r}} \hat{S}^{(a)}_{\ve{r}+\ve{e}_i}
   + J_{\text{pin}} \sum_{a} \hat{S}^{(a)}_{\ve{0}} \hat{S}^{(a)}_{\ve{0}+\ve{e}_{\rm x}},
   \label{Hpin}
\end{align}
with $J_{\text{pin}} > J$.
Therefore, we explicitly choose
one of multiple degenerate ground states by pinning the bond $(\ve{0}, \ve{0}+\ve{e}_{\rm x})$ and the bond observable 
\begin{align}
\label{eq:pinning_obs}
B_i(\ve{r}) \equiv \frac{1}{C(N,S)} 
  \Braket{ \sum_a \hat{S}^{(a)}_{\ve{r}} \hat{S}^{(a)}_{\ve{r}+\ve{e}_i}}
  \qquad i = {\rm x, y}
\end{align}
does not vanish as in the unpinned case.
  We notice that the AFM interactions in the Hamiltonian favors a minimization of $B_i(\ve{r})$.
  As proven in Appendix~\ref{app:bounds}, $B_i(\ve{r})\geq -1$ and approaches $-1$ when the two spins form a singlet.

We have chosen $J_{\text{pin}}$ such that the pinned bond satisfies  
\begin{multline}
\Braket{\sum_a \hat{S}^{(a)}_{\ve{0}} \hat{S}^{(a)}_{\ve{0}+\ve{e}_{\rm x}}}\\
\approx
\frac{1}{2} \left( \frac{1}{2 N_r}
\sum_{(\ve{r}, i), a} \Braket{\hat{S}^{(a)}_{\ve{r}} \hat{S}^{(a)}_{\ve{r}+\ve{e}_i}} - C(N,S)
\right).
\end{multline}
The  right-hand  side  corresponds  to  the point halfway between the background and the minimal value of
  $\Braket{\sum_a \hat{S}^{(a)}_{\ve{0}} \hat{S}^{(a)}_{\ve{0}+\ve{e}_{\rm x}}}$.  This results  in $J_{\text{pin}}$ between $1.2 J$ and $1.5 J$.

At large distances from the pinned bond, one will either be able to explicitly observe the ``selected'' order through $B_i(\ve{r})$ (if $\phi$ or $\psi$ are nonzero), or the order will vanish.

\subsection{$S=1/2$}
\label{sec:results:1_2}

\begin{figure}
  \centering
  \includegraphics[width=\linewidth]{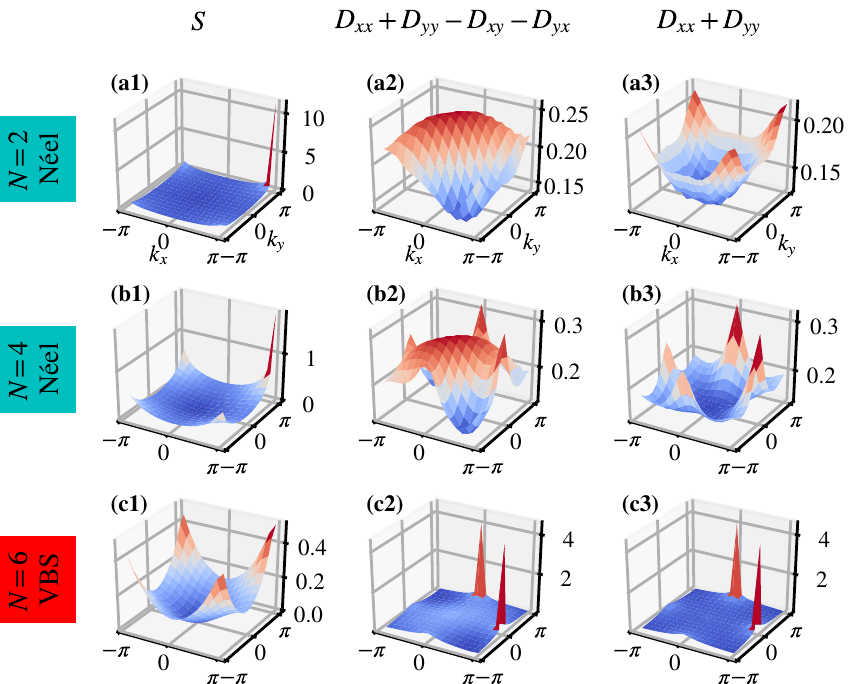}
  \caption{
  \label{fig:correlations_NS1}
  Correlation functions $S(\ve{k})$ [Eq. \eqref{Scorr}] and $D_{ij}(\ve{k})$ [Eq. \eqref{Scorr}]
  for the representation of Fig.~\ref{young} with $S=1/2$ and different values of $N$.}
\end{figure}
We first study the representations of Fig.~\ref{young} with $S=1/2$.
In Fig.~\ref{fig:correlations_NS1}, we show the structure factor $S(\ve{k})$ [Eq.~(\ref{Scorr})], and the two combinations of dimer correlations $D_{ij}(\ve{k})$ appearing in the definitions of the order parameter $\phi$ [Eq.~(\ref{phi})] and $\psi$ [Eq.~(\ref{psi})].
By considering three values of $N$,
we analyze the evolution of the order parameters across the transition between the N\'eel and VBS order.
For $N=2$, we realize a standard $S=1/2$ Heisenberg model on the square lattice, which displays a N\'eel order in the ground state.
As expected, $S(\ve{k})$ shows a strong peak at $(\pi,\pi)$; in comparison, the other order parameters are suppressed.
Upon increasing $N$ to $N=4$, we observe the emergence of peaks at $(\pi,0)$ and $(0,\pi)$ for the other order parameters.
As shown below using the correlation ratios, though close to the phase transition to the VBS state, the ground state is still N\'eel ordered.
For $N=6$ we observe a strong peak at $(\pi,0)$ and $(0,\pi)$ for the $D_{xx}(\ve{k}) + D_{yy}(\ve{k})$ order parameters, indicating the realization of the VBS phase.

\begin{figure}
  \centering
  \includegraphics[width=\linewidth]{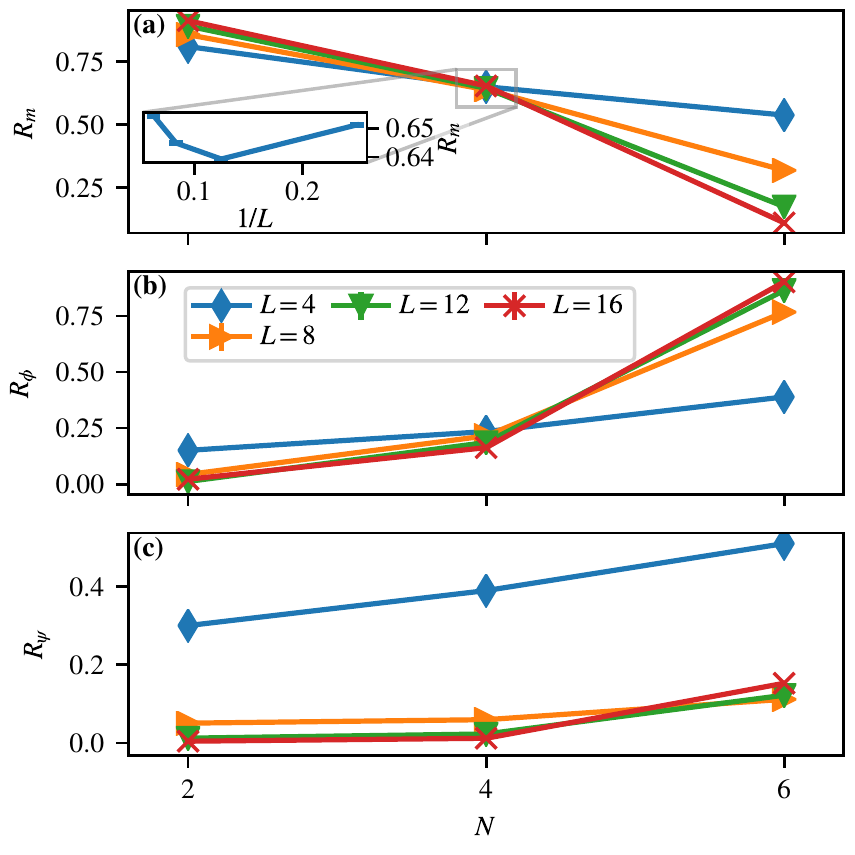}
  \caption{Correlation ratios of the staggered magnetization $m$ [Eq.~(\ref{staggm})], $\phi$ [Eq.~(\ref{phi})] and $\psi$ [Eq.~(\ref{psi})] order parameters for $S=1/2$, as a function of $N$, and for lattice sizes $L=4-16$.}
  \label{fig:results_NS1}
\end{figure}
To obtain a reliable determination of the ground state, we study the correlation ratios of the three order parameters discussed in Sec.~\ref{sec:results:op}.
In Fig.~\ref{fig:results_NS1}, we show the three correlation ratios $R_m$, $R_\phi$, and $R_\psi$ as a function  of $N$, for lattice sizes $L=4, 8, 12, 16$.
The crossing plot of $R_m$ indicates the disappearance of N\'eel order at $N\approx 4$.
At the same time, the curves of $R_\phi$ and $R_\psi$ exhibit a crossing for values of $N>4$; in particular, for $N=4$ both $R_\phi$ and $R_\psi$ decrease with the lattice size, indicating that both VBS and nematic order are short-ranged for $N=4$.
Therefore, as anticipated above, for $N=4$ the ground state is still a N\'eel order.   This  confirms  the result of Refs.~\cite{Paramekanti07,Kim_F18,WangD14}.
At $N=6$, $R_m$ decreases with $L$, while $R_\phi$ and $R_\psi$ increases in $L$, for $L\ge 6$; 
the $L=4$  data set  is  dominated  by   finite-size  effects.
Accordingly, for $N=6$ the ground state realizes VBS order.

\begin{figure}
  \centering
  \includegraphics[width=\linewidth]{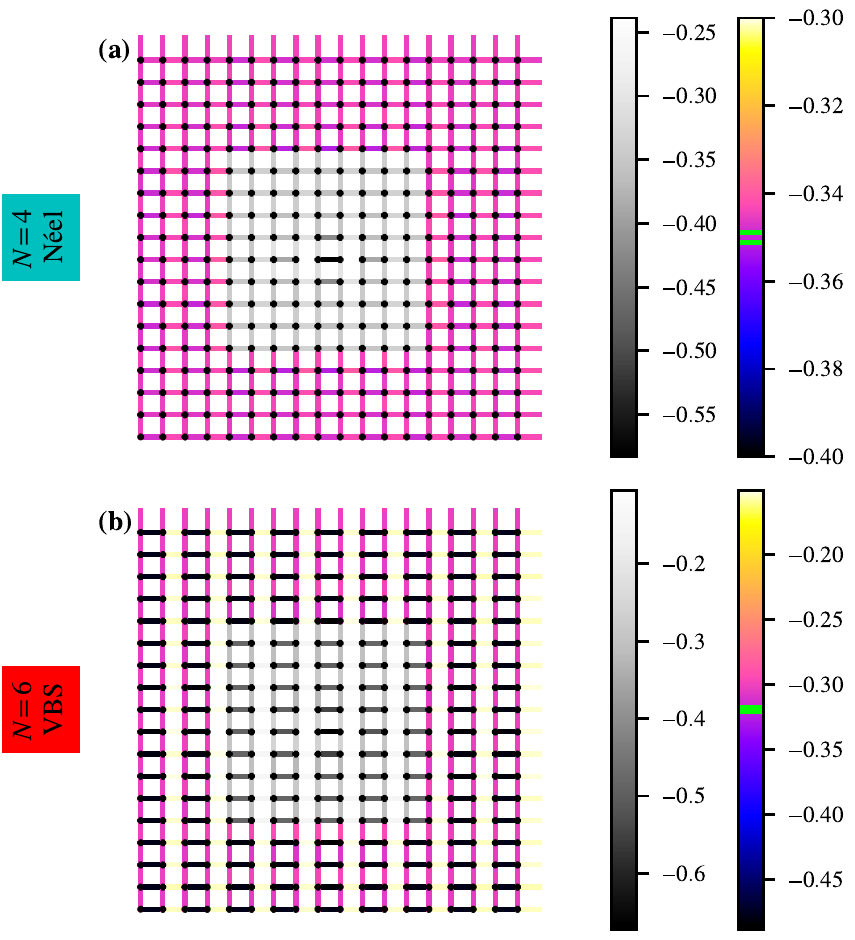}
  \caption{
  \label{fig:pin_NS1}
  Real-space value of bonds $B_i(\ve{r})$ [Eq.~\eqref{eq:pinning_obs}], as measured after pinning the central bond, for $S=1/2$ and lattice size $L=18$.
  Due to the observed different variations in the bond strength, in order to better highlight the patterns of bond correlations we have used different color scales for the region close and far from the pinned bond. The biggest error of the outer bonds is indicated by two red lines on the color scale.
  We have symmetrized the results with regards to inversions $y \rightarrow -y$, $x \rightarrow -x$ around the pinned bond.
  }
\end{figure}
These observations are confirmed by the real-space plot of the bond intensity shown in Fig.~\ref{fig:pin_NS1}, as obtained with the pinning-field method.
For $N=4$, in the vicinity of the pinned bond we observe a pattern reminiscent of the VBS order of Fig.~\ref{fig:vbs_nematic}.
However, at larger distances the modulation of bond intensity quickly decays, confirming that VBS order is actually short ranged.
For $N=6$, we observe a very clear pattern of strong and weak bonds that realize the VBS order.

\subsection{$S=1$}
\label{sec:results:1}

\begin{figure}
  \centering
  \includegraphics[width=\linewidth]{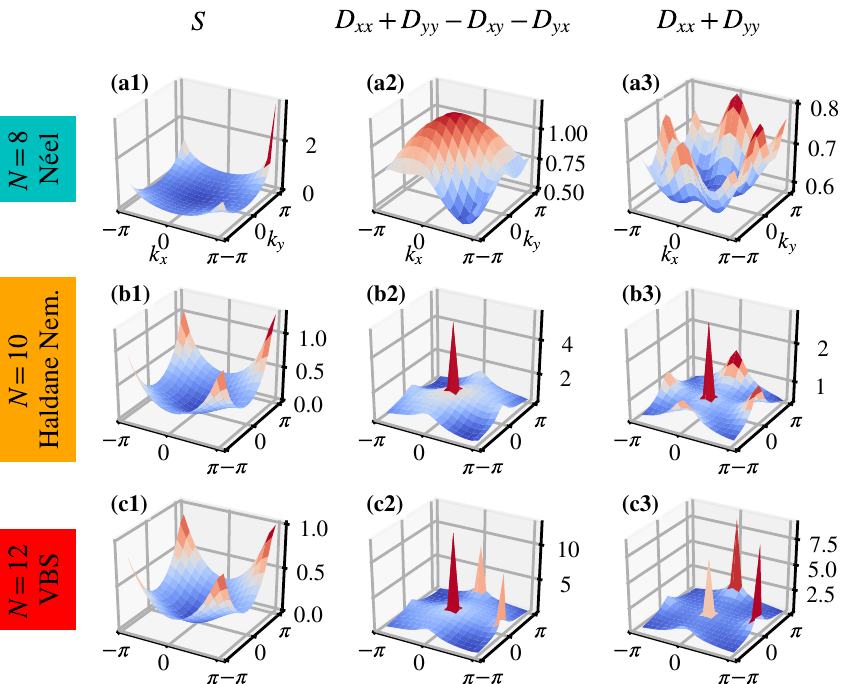}
  \caption{Same as Fig.~\ref{fig:correlations_NS1} for $S=1$}
  \label{fig:correlations_NS2}
\end{figure}
In studying the representations with $S=1$, we proceed analogously to the $S=1/2$ case discussed in Sec.~\ref{sec:results:1_2}.
In Fig.~\ref{fig:correlations_NS2}, we show the order parameters in momentum space for the case $S=1$, and three representative values of $N$.
For $N=8$, a clear peak of the spin structure factor at $(\pi, \pi)$, along with a comparatively smoother momentum dependence of the other order parameters, indicate the presence of N\'eel order.
At $N=10$, we observe instead the emergence of a clear signal of the nematic order parameter at zero momentum.
The VBS order parameter exhibits a similar peak at zero momentum, whose signal predominantly arises from the nematic order parameter, while at momentum $(\pi,0)$ a subdominant peak is observed.
The N\'eel order parameter instead does not show a predominant signal at $(\pi,\pi)$, but rather equally large values at 
the corners of the Brillouin zone.
These behaviors suggest the onset of the two-fold degenerate nematic order.
For $N=12$ a clear signal at $(\pi,0)$ momentum appears in the VBS order parameter.
Along with the sharp zero-momentum value of the nematic order parameter, these findings suggest the realization of VBS order for $N=12$.

\begin{figure}
  \centering
  \includegraphics[width=\linewidth]{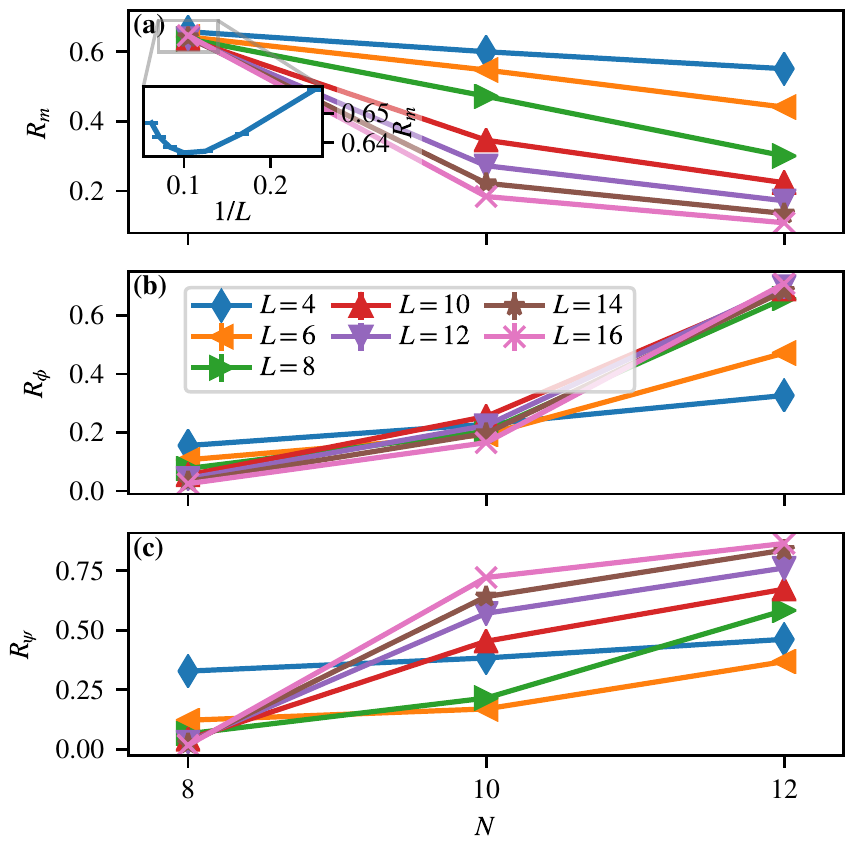}
  \caption{
  \label{fig:results_NS2}
  Same as Fig.~\ref{fig:results_NS1} for $S=1$.
  }
\end{figure}
As we did for the $S=1/2$ case, the above qualitative observations on the momentum dependence of the various order parameters can be put on firm ground by examining the correlation ratios shown in Fig.~\ref{fig:results_NS2}.
The magnetic correlation ratio $R_m$ displays a crossing at about $N\approx 8$, such that for $N>8$, $R_m$ decreases with the lattice size.
On the other hand, at $N=8$ both $R_\phi$ and $R_\psi$ decrease on increasing $L$, implying that both VBS and nematic order are short ranged.
Therefore, one can conclude that for $N=8$ the ground state is antiferromagnetically ordered.
The behavior of $R_\phi$ shows rather important finite-size corrections.
In fact, while curves for $L\le 10$ cross for $8 < N < 10$, the crossing point quickly increases with $L$, such that for $L\ge 12$ a crossing is found for $10 < N <12$.
In particular, $N=10$ has a nonmonotonic behavior, increasing in $L$ for $L\le 10$, and decreasing for $L\ge 12$.
This observation supports the presence of a significant, but still short-ranged, VBS order, which is responsible for important finite-size corrections.
The nematic correlation ratio $R_\psi$ shows a crossing between $N=8$ and $N=10$.
Also, here we observe a clear drift in the crossing of $R_\psi$, although the situation for $N=10$ is rather clear and indicates long-range order in $\psi$.
In view of these observations, and
referring to Table \ref{order_parameters}, we conclude that a nematic ground state is realized for $N=10$, while for $N\ge 12$ the ground-state is VBS ordered.

\begin{figure}
  \centering
  \includegraphics[width=\linewidth]{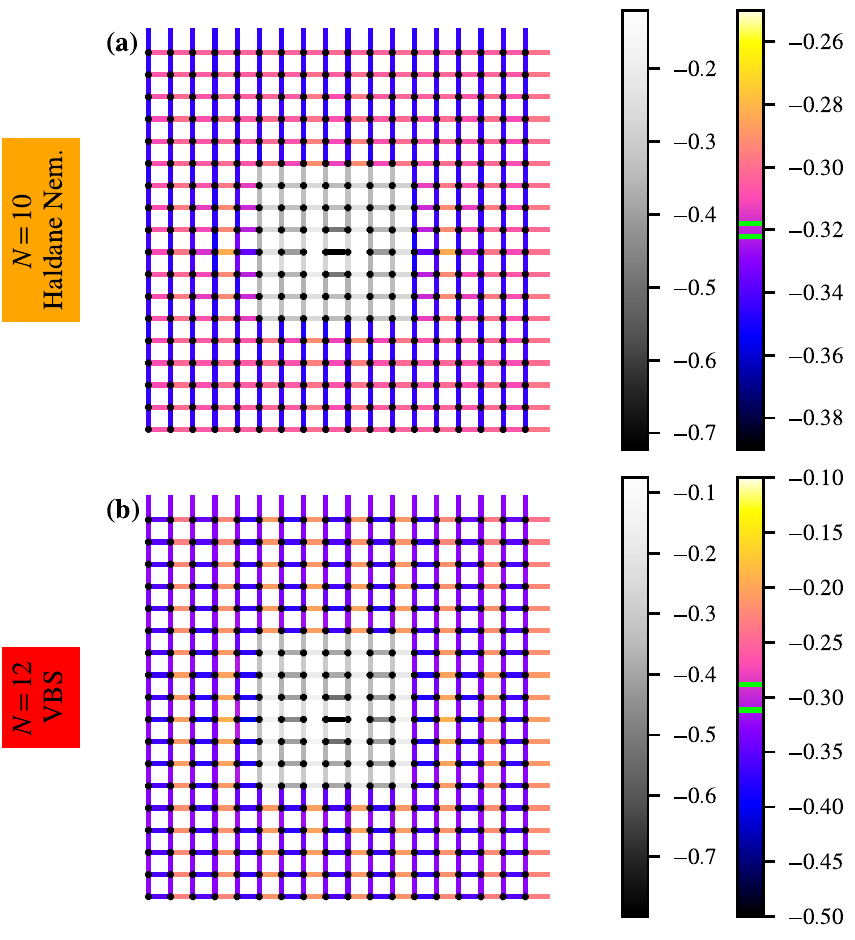}
  \caption{
  \label{fig:pin_NS2}
  Same as Fig.~\ref{fig:pin_NS1} for $S=1$.
  }
\end{figure}
These conclusions are nicely confirmed by the real-space plots of the bond strength obtained with the pinning-field method and shown in Fig.~\ref{fig:pin_NS2}.
For $N=10$ we clearly observe the formation of a twofold degenerate stripe-like structure, signalling the presence of the nematic order.
For $N=12$ instead a VBS order is found.
Interestingly, while for $S=1/2$ the VBS order found at $N=6$ (Fig.~\ref{fig:pin_NS1}) resembles the ladder order illustrated in Fig.~\ref{fig:vbs_nematic}, for $S=1$ and $N=12$ the VBS pattern shown in Fig.~\ref{fig:pin_NS2} rather suggests the plaquette order of Fig.~\ref{fig:vbs_nematic}.

\subsection{$S=3/2$}
\label{sec:results:3_2}

\begin{figure}
  \centering
  \includegraphics[width=\linewidth]{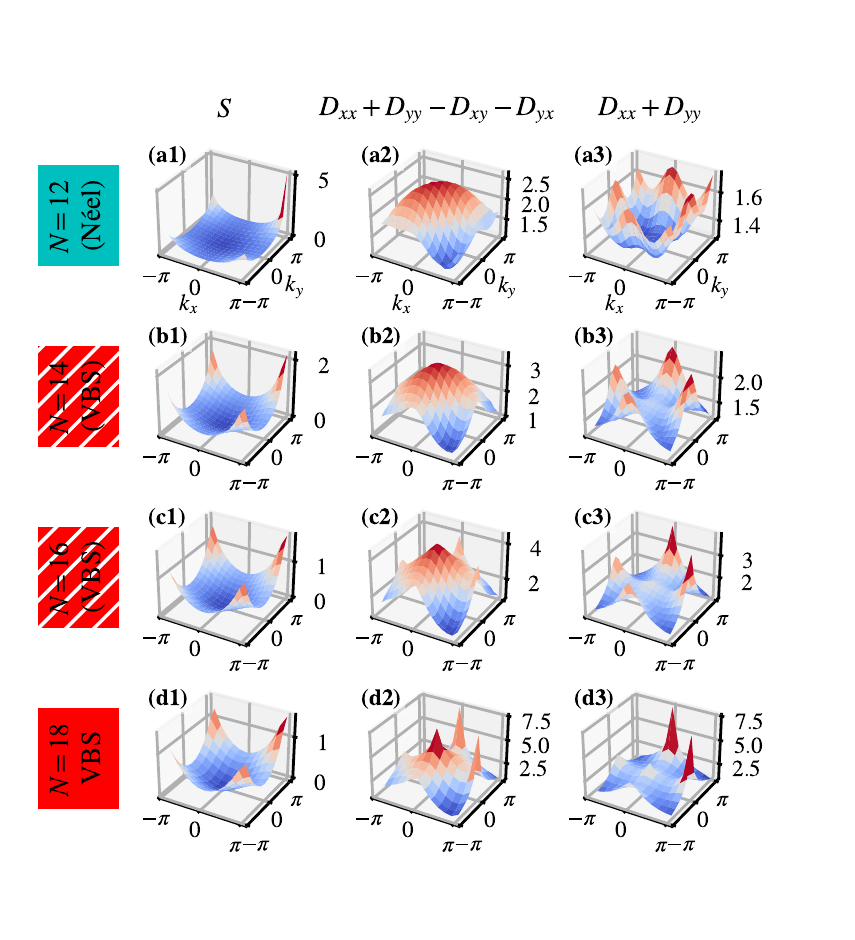}
  \caption{Same as Fig.~\ref{fig:correlations_NS1} for $S=3/2$}
  \label{fig:correlations_NS3}
\end{figure}
In Fig.~\ref{fig:correlations_NS3}, we show the order parameters in momentum space, and $N=12-18$.
Analogous to the cases analyzed in the previous sections, for $N=12$ we find a signal of N\'eel order.
In the region $14\le N\le 16$, QMC data do not allow us to unambiguously single out the ground state.
Upon  increasing $N$, the peak at $(\pi,\pi)$ in $S(\ve{k})$ slowly decreases in magnitude.
At the same time, we observe the appearance of a maximum in the nematic order parameter at zero momentum, and in the VBS order parameter for $(\pi, 0)$ and $(0,\pi)$ momenta.
Eventually, for $N=18$ the momentum structure of the order parameters more clearly favors the realization of VBS order.
The observed behavior suggests a comparatively  broad  critical region around $14\le N\le 18$.

\begin{figure}
  \centering
  \includegraphics[width=\linewidth]{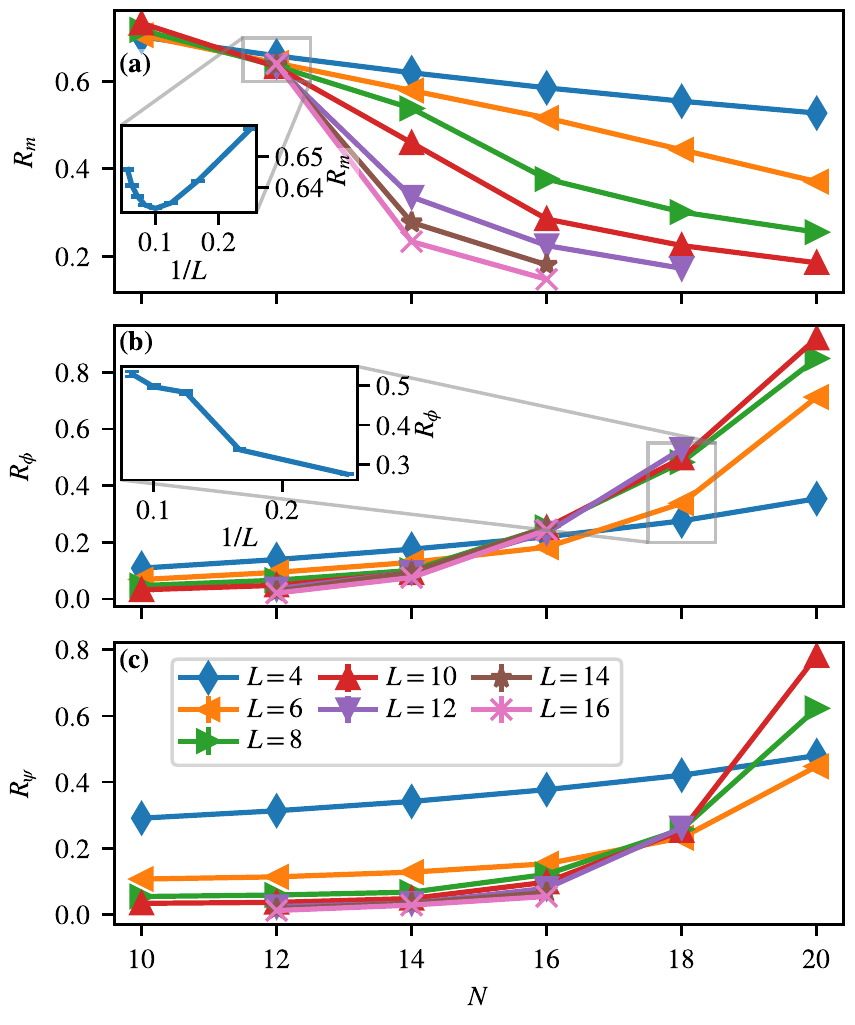}
  \caption{
  \label{fig:results_NS3}
  Same as Fig.~\ref{fig:results_NS1} for $S=3/2$}
\end{figure}
In an attempt to better understand the ground-state diagram for $S=3/2$, as for the other values of $S$ we have analyzed the correlation ratios, shown in Fig.~\ref{fig:results_NS3}.
Due to the increased computational costs,
 we restricted the simulations for the larger lattice sizes $L\ge 12$ to the more involved cases $12\le N\le 16$.
For smaller lattice sizes $L\le 10$, the curves for $R_m$ appear to cross at a value of $N$ very close, but smaller than $N<12$.
We observe, however, some drift towards larger values of $N$ in the crossings.
Furthermore, as shown in the inset of Fig.~\ref{fig:results_NS3}, $R_m$ at $N=12$ exhibits an upwards trend for $L>10$.
Together with the observed slow
decrease of $R_\phi$ and $R_\psi$ in $L$ for $N=12$, this implies N\'eel order for $N=12$.

For $N\ge 14$, the $L-$dependence of $R_m$ clearly rules out  N\'eel order.
On the other hand, $R_\phi$ slowly decreases with $L$ for $N=14$, and for $N=16$ it grows slightly
up to $L=8$. In both cases, QMC data for $L\ge 8$ are indistinguishable within error bars.
A similar flattening of QMC data is found in $R_\psi$ for $N=14, 16$.
This behavior does not allow us to draw firm conclusions on the nature of the ground state for $N=14,16$.
Since a N\'eel state can be ruled out,
a reasonable hypothesis is the realization of a VBS state, however, with a weak order parameter.

For $N>16$, both $R_\phi$ and $R_\psi$ show a crossing close to $N=18$. Furthermore, we observe a monotonic growth of $R_\phi$ in $L$ for $N=18$.
This leads us to conclude a VBS order for $N=18$.

\begin{figure}
  \centering
  \includegraphics[width=\linewidth]{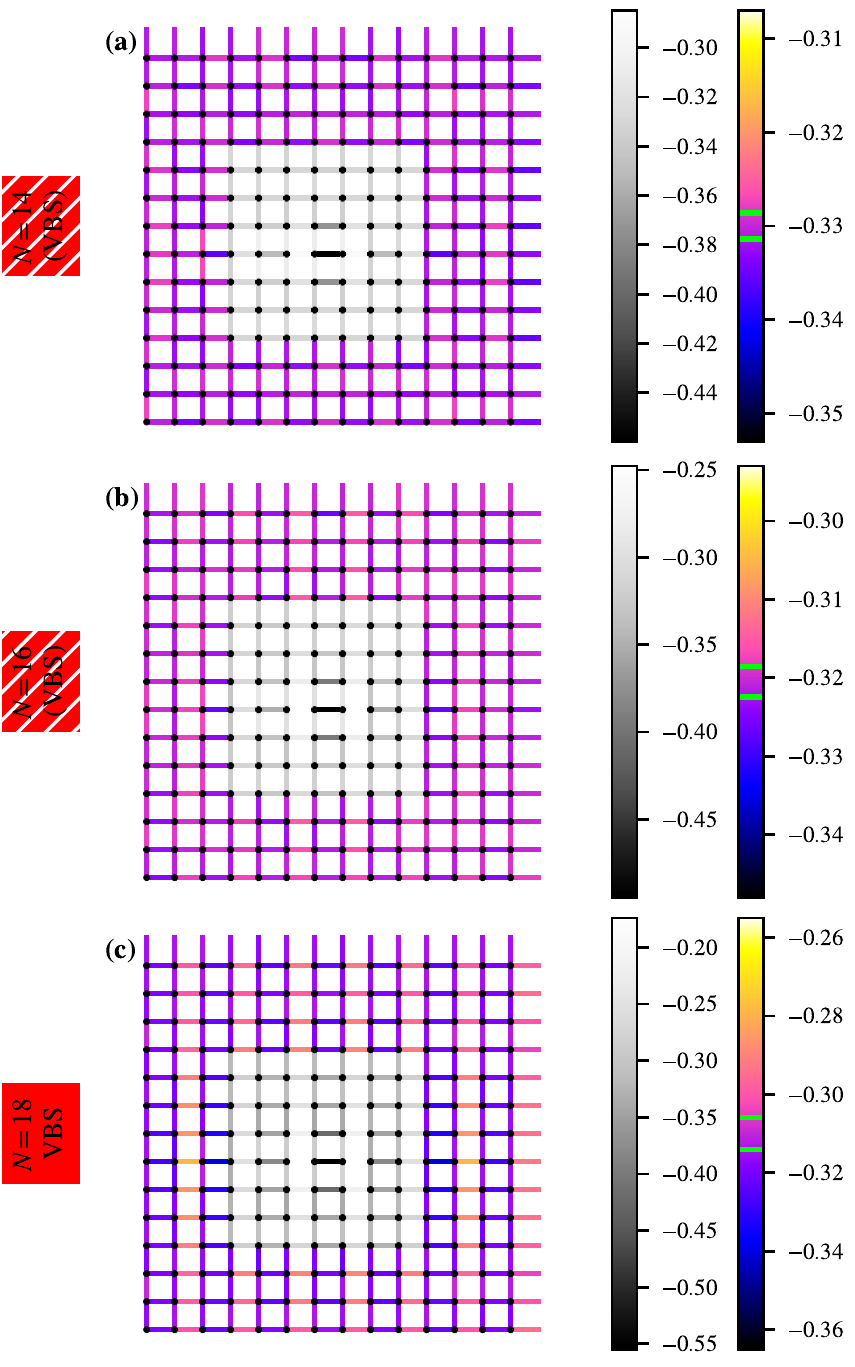}
  \caption{
  \label{fig:pin_NS3}
  Same as Fig.~\ref{fig:pin_NS1} for $S=3/2$ and lattice size $L=14$ $(N=14,16)$, $L=12$ $(N=18)$.
  }
\end{figure}
Finally, we have studied the pattern of bond strength in real space with the pinning-field method.
The results for $N=14,16,18$ are reported in Fig.~\ref{fig:pin_NS3}.
For $N=14,16$, despite some signs of dimerization, we do not observe a clear VBS pattern.
In line with the previous analysis,
for $N=18$ we find a bond dimerization which confirms a VBS ground state.

\subsection{$S=2$}
\label{sec:results:2}

\begin{figure}
  \centering
  \includegraphics[width=\linewidth]{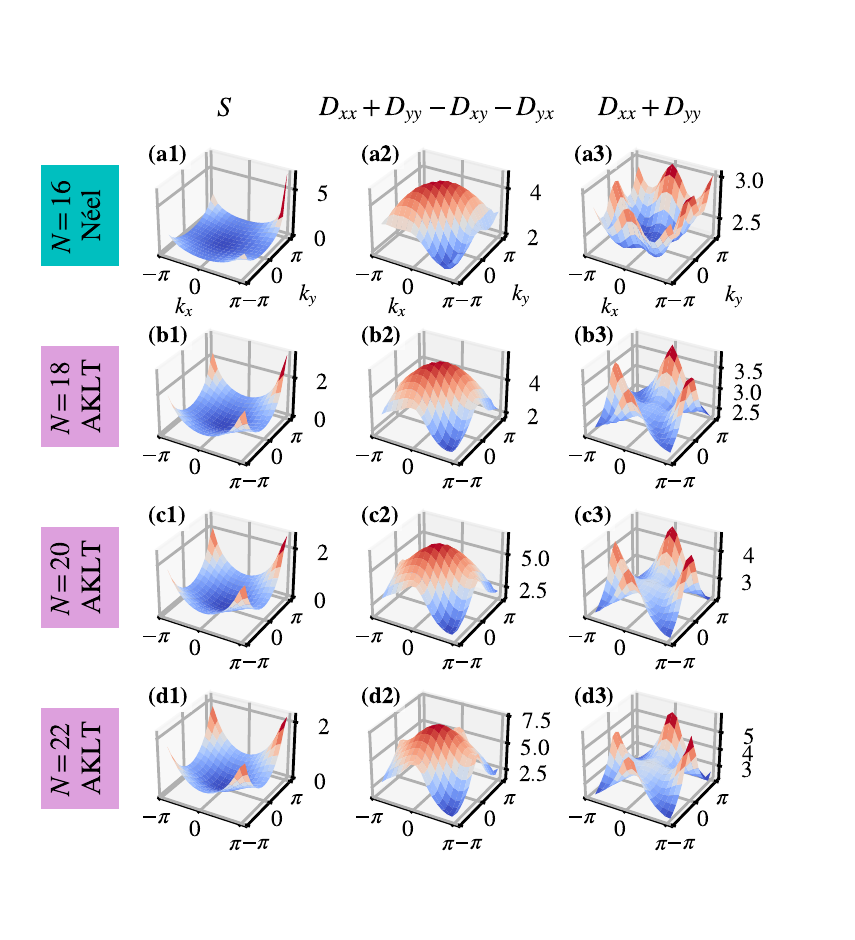}
  \caption{Same as Fig.~\ref{fig:correlations_NS1} for $S=2$}
  \label{fig:correlations_NS4}
\end{figure}

\begin{figure}
  \centering
  \includegraphics[width=\linewidth]{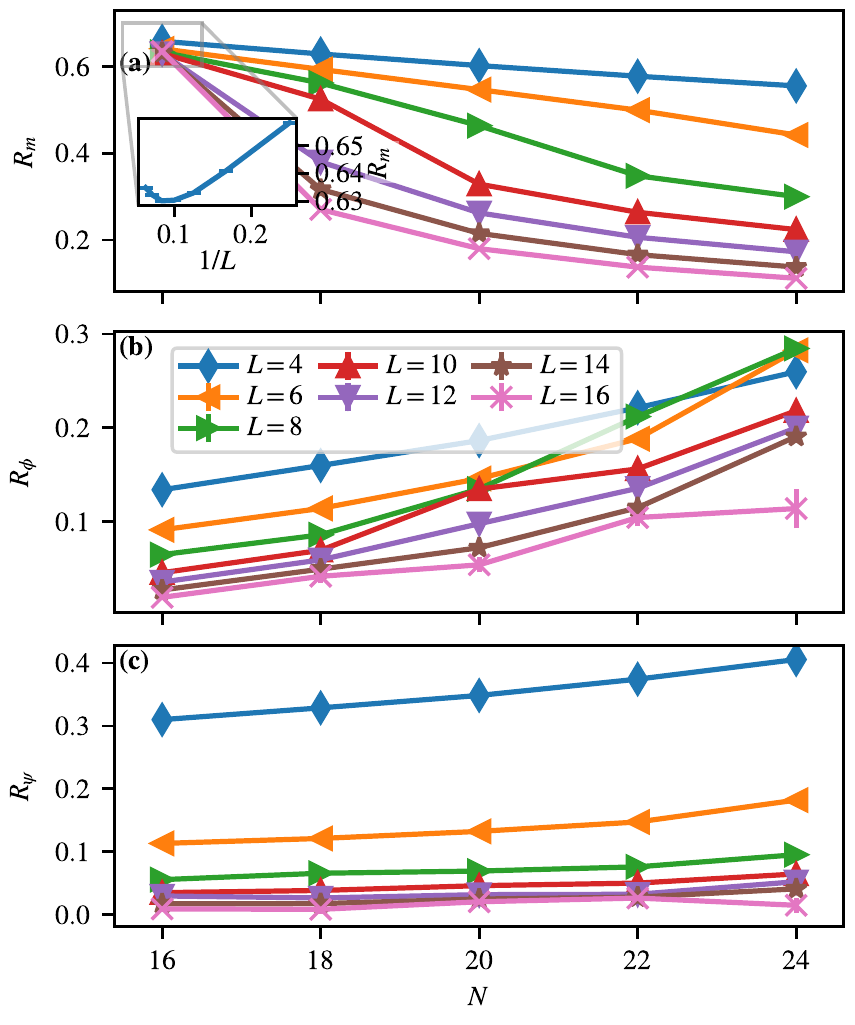}
  \caption{
  \label{fig:results_NS4}
  Same as Fig.~\ref{fig:results_NS1} for $S=2$}
\end{figure}

As for previous values of $S$, we begin our investigation for $S=2$ with momentum space plots of correlation functions shown in Fig.~\ref{fig:correlations_NS4}. At $N=16$, the spin structure factor shows a sharp peak at $(\pi,\pi)$, indicating long-range AFM order. For bigger values of $N$ the peak weakens and broadens, suggesting short-range AFM order. The correlations for nematic and VBS order show only very broad maxima for the range $N \in [16, 22]$, implying the absence of both of these  orders. The correlation ratios plotted in Fig.~\ref{fig:results_NS4} support these qualitative observations. The inset in Fig.~\ref{fig:results_NS4}(a) shows that while $R_m$ decreases at $N=16$ from $L=4$ to $L=12$, the trend is reversed on bigger lattices and $R_m$ increases from $L=12$ to $L=16$. This indicates that $N=16$ has a Néel ground state which is close to a competing order. Both $R_\phi$ and $R_\psi$ decrease with increasing system size in the investigated range $N \in [16, 24]$. As per Table~\ref{order_parameters}, this leaves a two-dimensional AKLT order as a ground-state candidate for $N \in [18, 24]$. With the AKLT   construction corresponding  to the right-hand side of Fig.~\ref{fig:vbs_nematic}(e), we  understand that each boundary  site   hosts an \suna\ representation corresponding  to one  column ($S=1/2$)   and   $N/2$  rows.  Hence  the  boundary   defines  a one-dimensional chain in the aforementioned representation.    It is known  that for  $N \geq 4$  this chain  dimerizes \cite{Onufriev99,Assaraf04,Paramekanti07,Kim_F18}.

\begin{figure}
  \centering
  \includegraphics[width=\linewidth]{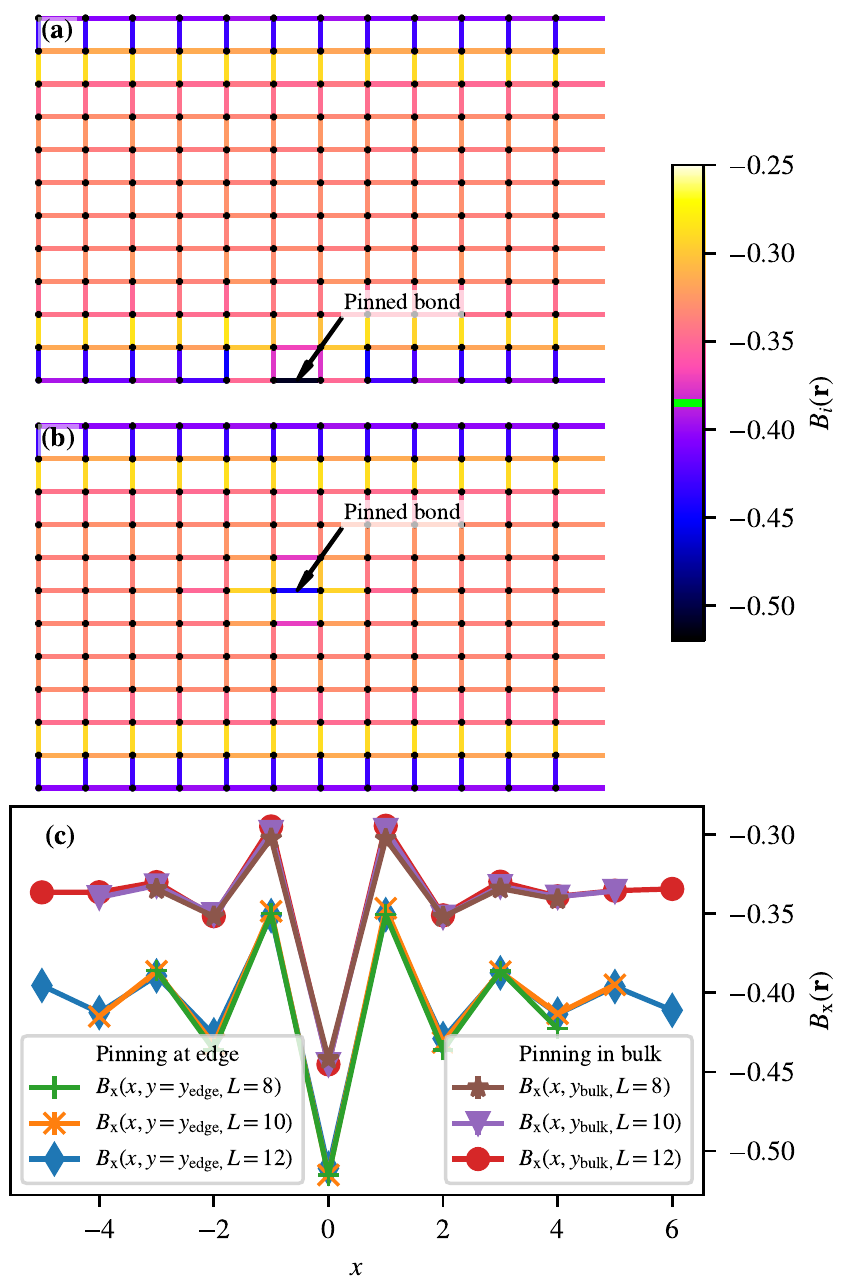}
  \caption{
  \label{fig:pinning_NS4_N18}
  Real-space value of bonds $B_i(\ve{r})$ for $S=2$, $N=18$, lattice size $L \in \{8, 10 , 12\}$ and open boundary conditions in $y$ direction. Comparison between pinning a bond at the edge \textbf{(a)} and a bond in the bulk \textbf{(b)}. 
  \textbf{(c)} $B_{\rm x}(\ve{r})$ on horizontal lines through the pinned bonds.
  }
\end{figure}

To further investigate this possibility, we simulate the model
on a lattice with periodic boundary conditions in the $x$ direction and open boundary conditions  along $y$, corresponding to a cylinder geometry. Fig.~\ref{fig:pinning_NS4_N18} shows the results for $S=2,N=18$ in a pinning-field approach with pinned bonds at the edge and in the bulk, respectively. The induced dimerization pattern propagates on the edge much further than in the bulk, supporting the presence of an AKLT phase  with  
boundary corresponding to an $S=1/2$ \sun\ chain  in the  totally  antisymmetric  self-adjoint representation.

\section{Summary}
\label{sec:summary}
We have studied the ground-state phase diagram of an \sun\ AFM model on the square lattice,    with   irreducible representations 
of \suna\ illustrated in Fig.~\ref{young} and characterized by  a  Young tableau  consisting  of  $2S$ columns  and  $N/2$  rows. 
For  even  values of  $N$   we   have presented    negative  sign free  QMC   data  that  results  in  the   rich phase  diagram  of  
Fig.~\ref{fig:PhaseDiagram}. 
In line with field-theoretical studies \cite{RS-89,RS-89b,RS-90},
for any value of the generalized spin $S$, we found N\'eel order at small values of $N$, and a dimerized VBS state for large $N$.     The   disordered  
 states  proximate  to the melting of  the  N\'eel  state    can be naturally understood  in terms of  condensation of  monopoles.  
 These  states  turn out  to be located    along  the  line  $N=8S +  2 $ in the  $S$  versus $N$  phase  diagram. 
   At  $S=1$  we  observe   nematic  AKLT \cite{AKLT-88}   states     where    $C_4 $ symmetry    is  spontaneously  reduced  to $C_2$,    with the emergence of spin-1 chains  along  one lattice  direction.      At $S=2$,  the   AKLT   construction  provides   an understanding of  the  non-degenerate  state.  In fact,  this  construction  can  be  generalized  to any  spin  $S=Z/2$  system   on  a  lattice  with  coordination number $Z$; an example  for the   honeycomb lattice    is  given in Refs.~\cite{Pomata20,Poilblanc13}. In our  specific  case,  the  edge  state    corresponds  to  an \sun\ spin system in an irreducible representation specified by a  Young  tableau with one  column ($S=1/2$)  and  $N/2$   rows.   At  $N \geq 4$, such a state is know  to dimerize   \cite{Kim_F18}. Our simulations with open  boundary  conditions  support  this picture.      For  half-integer  values of $S$,   a  four  fold  degenerate  VBS  state 
   emerges.  The  detailed   nature of  the   dimerization  was  studied  using   a pinning-field  approach \cite{AH-13}. 

While the resulting phase diagram reproduced in Fig.~\ref{fig:PhaseDiagram}  will serve as a benchmark for future studies, our findings points to future avenues  of  research.
It would be very interesting to study in detail the quantum phase transitions between the various states.
Although the present setup allows us to consider integer values of $N$ only, it may be possible to investigate the phase transitions with a suitably  defined designer Hamiltonian containing, e.g., some interactions that favor a specific phase, so as to be able to interpolate between them.     The  topological  arguments  that  lead  to   the observed  phase  diagram carry  over  to  other  representation of \suna\, such that further  calculations with  alternative   methods   such as  stochastic series expansion \cite{Kaul12}  are  certainly   desirable. 

\begin{acknowledgments}
We  would  like to thank  S.~Sachdev  for  illuminating discussions.
This research has been funded by the Deutsche Forschungsgemeinschaft (DFG) through the W\"urzburg-Dresden Cluster of Excellence on Complexity and Topology in Quantum Matter \textit{ct.qmat} -- Project No.~390858490 (F.\,F.\,A.),  the SFB~1170 on Topological and Correlated Electronics at Surfaces and Interfaces -- Project No.~258499086 (J.\,S., F.\,F.\,A.), Project No. 414456783 (F.\,P.\,T.) and Grant No.~AS~120/14-1 (F.\,F.\,A.).
FFA acknowledges financial support from Deutsche Forschungsgemeinschaft under the grant AS 120/16-1 (Project number 493886309) that is part of the collaborative research unit on \textit{Correlated Quantum Materials \& Solid State Quantum Systems}.
  The authors gratefully acknowledge the Gauss Centre for Supercomputing e.V. (\href{www.gauss-centre.eu}{www.gauss-centre.eu}) for funding this project by providing computing time on the GCS Supercomputer SuperMUC-NG at Leibniz Supercomputing Centre (\href{www.lrz.de}{www.lrz.de}), where part of the simulations have been carried out.
  The authors gratefully acknowledge the scientific support and HPC resources provided by the Erlangen National High Performance Computing Center (NHR@FAU) of the Friedrich-Alexander-Universität Erlangen-Nürnberg (FAU) under the NHR project b133ae. NHR funding is provided by federal and Bavarian state authorities. NHR@FAU hardware is partially funded by the German Research Foundation (DFG) – 440719683.
\end{acknowledgments}

\appendix
\section{The quadratic Casimir eigenvalue in terms of the Young tableau}
\label{app:casimir_young}
In this appendix, we discuss the relation between the Young tableau of an irreducible representation and the corresponding eigenvalue of the quadratic Casimir operator.
For an irreducible representation, whose Young tableau has $n_l$ rows of length $\{l_i\}$ and $n_c$ columns of length $\{c_i\}$, the eigenvalue of the quadratic Casimir operator is \cite{PS-84}
\begin{equation}
  C = \frac{1}{2} \left[ r\left(N-\frac{r}{N}\right)+\sum_i^{n_l}l_i^2-\sum_i^{n_c}c_i^2\right],
  \label{casimir_young}
\end{equation}
where $r=\sum_i l_i=\sum_i c_i$ is the total number of boxes.

In this appendix, we also derive Eq.~(\ref{casimir_young}), which is stated in the Appendix of Ref.~\cite{PS-84} without an explicit proof.
First, we notice that in Eq.~(\ref{casimir_young}) there is an implicit choice of normalization. As we show below, such a normalization is consistent with Eq.~(\ref{casimir_def}).

For an irreducible representation, the value of $C$ can be easily computed with Weyl's formula \cite{Vergados_book,note-weyl},
\begin{equation}
  C = \langle\Lambda|\Lambda + 2\delta\rangle,
  \label{weyl}
\end{equation}
where $\Lambda$ is the maximum weight of the representation and $\delta$ the Weyl vector. In the Dynkin representation the metric tensor of the scalar product is, up to a normalization $\cal N$, the inverse of the transpose of the Cartan matrix $A$ \cite{Vergados_book},
\begin{equation}
  \begin{split}
    G^{ij} &= {\cal N}\left[\left(A^T\right)^{-1}\right]_{ij},\\
    \left[\left(A^T\right)^{-1}\right]_{ij} &= \text{min}(i,j)-\frac{ij}{N},
  \end{split}
  \label{metric_dynkin}
\end{equation}
and the Weyl vector is $\delta=(1,1,\ldots, 1)$.
To fix the normalization ${\cal N}$, we compute $C$ for the defining representation, and match it with Eq.~(\ref{casimir_def}). For the defining representation, the maximum Dynkin weight is $\Lambda_{\alpha_i}=\delta_{i,1}$, hence
\begin{equation}
  \begin{split}
    C &= {\cal N} \sum_{i,j=1}^{N-1}\delta_{i,1}\left[\text{min}(i,j)-\frac{ij}{N}\right]\left(\delta_{j,1} + 2\right)\\
    &= {\cal N}\frac{N^2-1}{N}.
  \end{split}
  \label{casimir_defrep}
\end{equation}
On the other hand, by taking the trace on both hand sides of Eq.~(\ref{casimir_def}), $C$ is readily computed as $C=(N^2-1)/(2N)$. Therefore the normalization constant is ${\cal N}=1/2$.

Employing Eq.~(\ref{weyl}), we first compute $\langle \Lambda|\Lambda\rangle$. Using Eq.~(\ref{young_to_dynkin})
\begin{equation}
  \langle \Lambda|\Lambda\rangle = \sum_{i,j=1}^{N-1}\left(l_i-l_{i+1}\right)G^{ij}\left(l_j-l_{j+1}\right),
  \label{LL_young}
\end{equation}
where the metric tensor $G^{ij}$ is given in Eq.~(\ref{metric_dynkin}). By developing the products and employing change of variables $i\rightarrow i-1$, $j\rightarrow j-1$, Eq.~(\ref{LL_young}) can be written as
\begin{equation}
  \begin{split}
    \langle \Lambda|\Lambda\rangle =& l_1G^{11}l_1\\
    &+l_1\sum_{j=2}^{N-1}\left(G^{1,j}-G^{1,j-1}\right)l_j\\
    &+\sum_{i=2}^{N-1}l_i\left(G^{i,1}-G^{i-1,1}\right)l_1\\
    &+\sum_{i,j=2}^{N-1}l_i\left(G^{i,j}-G^{i-1,j}-G^{i,j-1}+G^{i-1,j-1}\right)l_j,
  \end{split}
  \label{LL_young2}
\end{equation}
where we have used that $l_N=0$ for Young tableaux of \suna\ representations.
Using Eq.~(\ref{metric_dynkin}), we have
\begin{equation}
  G^{1,j}-G^{1,j-1} = G^{i,1}-G^{i-1,1} = -\frac{1}{2N},
  \label{Gdiff}
\end{equation}
where we have employed the normalization ${\cal N}=1/2$ obtained after Eq.~(\ref{casimir_defrep}).
Further, using Eq.~(\ref{metric_dynkin}), the difference in the parenthesis in the last term of Eq.~(\ref{LL_young2}) is computed as
\begin{equation}
  \begin{split}
    &G^{i,j}-G^{i-1,j}-G^{i,j-1}+G^{i-1,j-1} = \\
    &\frac{1}{2}\Big[\text{min}(i,j)-\text{min}(i-1,j)\\
      &-\text{min}(i,j-1)-\text{min}(i-1,j-1)-\frac{1}{N}\Big].
  \end{split}
  \label{Gdiff4}
\end{equation}
By enumerating the various cases, it is easy to see that
\begin{equation}
  \text{min}(i,j)-\text{min}(i-1,j)-\text{min}(i,j-1)-\text{min}(i-1,j-1)=\delta_{ij}.
  \label{mindiff}
\end{equation}
Using Eqs.~(\ref{Gdiff}), (\ref{Gdiff4}) and (\ref{mindiff}) in Eq.~(\ref{LL_young2}), we obtain the first term in Weyl's formula,
\begin{equation}
  \begin{split}
    \langle \Lambda|\Lambda\rangle =& \frac{l_1^2}{2}\left(1-\frac{1}{N}\right) -\frac{1}{N}l_1\left(r-l_1\right)\\
    &+\frac{1}{2}\sum_{i=1}^{N-1}l_i^2-\frac{1}{2}l_1^2+\frac{1}{2N}\left(r-l_1\right)^2\\
    =&\frac{1}{2}\left(\sum_{i=1}^{n_l}l_i^2-\frac{r^2}{N}\right),
  \end{split}
  \label{LL_young3}
\end{equation}
where $r=\sum_i l_i$ is the total number of boxes and the sum over $l_i^2$ can be restricted to the $n_l$ nonzero row lengths.

To compute the second term $\langle \Lambda|2\delta\rangle$ in Weyl's formula, we use a different parametrization of $\Lambda$. Since in the Dynkin representation the components of $\Lambda_{\alpha_i}$ are positive integers [see Eq.~(\ref{young_to_dynkin})], we can parametrize $\Lambda_{\alpha_i}$ as
\begin{equation}
  \Lambda_{\alpha_i} = \sum_{a=1}^{n_c}\delta_{i,c_a}.
  \label{column_to_dynkin}
\end{equation}
The set $\{c_a\}$ represents the position of the rows in the corresponding Young tableau where the number of boxes decreases on the following row. Such a decrease corresponds to the end of the column, hence $\{c_a\}$ are the column lengths. Using Eqs.~(\ref{column_to_dynkin}) and (\ref{metric_dynkin}) we have
\begin{equation}
  \begin{split}
    \langle\Lambda | 2\delta\rangle &=\frac{1}{2}\sum_{i,j=1}^{N-1}\sum_{a,b=1}^{n_c}\left(\delta_{i,c_a}\text{min}(i,j)2 - \delta_{i,c_a} \frac{ij}{N}2\right)\\
    &=\sum_{a=1}^{n_c}\sum_{j=1}^{N-1}\text{min}(c_a,j)-\sum_{a=1}^{n_c}\sum_{j=1}^{N-1}\frac{c_aj}{N}
  \end{split}
  \label{Ldelta_young}
\end{equation}
The first sum in Eq.~(\ref{Ldelta_young}) can be written as
\begin{equation}
  \begin{split}
    \sum_{a=1}^{n_c}\sum_{j=1}^{N-1}\text{min}(c_a,j) &= \sum_{a=1}^{n_c}\left(\sum_{j=1}^{c_a}j + \sum_{j=c_a+1}^{N-1}c_a\right)\\
    &=\left(N-\frac{1}{2}\right)r-\frac{1}{2}\sum_{a=1}^{n_c}c_a^2,
  \end{split}
  \label{Ldelta_young2}
\end{equation}
where we have used $\sum_a c_a=r$.
The second sum in Eq.~(\ref{Ldelta_young}) can be computed as
\begin{equation}
  \sum_{a=1}^{n_c}\sum_{j=1}^{N-1}\frac{c_aj}{N} = \frac{1}{2}r(N-1)
  \label{Ldelta_young3}
\end{equation}
Inserting Eqs.~(\ref{Ldelta_young2}) and (\ref{Ldelta_young3}) in Eq.~(\ref{Ldelta_young}), we obtain the second term of Weyl's formula
\begin{equation}
  \langle\Lambda | 2\delta\rangle = \frac{1}{2}\left(rN-\sum_{a=1}^{n_c}c_a^2\right)
  \label{Ldelta_young4}.
\end{equation}
Finally, employing Eqs.~(\ref{LL_young3}) and (\ref{Ldelta_young4}) in Eq.~(\ref{weyl}) one obtains Eq.~(\ref{casimir_young}).

As is known from the rules of Young tableaux of \suna\ representations, columns of length $N$ can be deleted since they correspond to an invariant under \sun. This is consistent with the formula of Eq.~(\ref{casimir_young}). Indeed, by adding to a Young tableau a column of length $N$, we have
\begin{equation}
  \begin{split}
    r&\rightarrow r + N,\\
    l_i&\rightarrow l_i+1,\\
    \sum_{i=1}^{n_c} c_i^2 &\rightarrow  N^2 + \sum_{i=1}^{n_c} c_i^2.
  \end{split}
\label{column_addition}
\end{equation}
Inserting the substitutions of Eq.~(\ref{column_addition}) in Eq.~(\ref{casimir_young}), one can check that the Casimir eigenvalue is left unchanged.

Finally, it is easy to check that in the case of the defining representation, whose Young tableau is a single box, Eq.~(\ref{casimir_young}) gives the expected result, with the normalization consistent with Eq.~(\ref{casimir_def}).

\section{Bound on the eigenvalue of the quadratic Casimir operator}
\label{app:casimir_max}
The tensor product of $2S$ self-adjoint antisymmetric representations given in Eq.~(\ref{antisym_rep}) decomposes into different irreducible representations.
In this appendix we prove that among those representations, the maximally symmetric one of Fig.~\ref{young} has the maximum Casimir eigenvalue, which we compute.

Due to the rules for the composition of Young tableaux, each of the irreducible representations arising from the tensor product has a Young tableau whose total number of boxes is $r\le(2S)(N/2)=NS$ and whose row lengths cannot exceed $2S$, $l_i\le 2S$. Thus an upper bound for $\sum_i l_i^2$ appearing in Eq.~(\ref{casimir_young}) is
\begin{equation}
  \sum_i^{n_l}l_i^2 \le \sum_i^{n_l}2Sl_i = 2Sr.
  \label{upper_bound_rows}
\end{equation}
This bound is saturated by
\begin{equation}
  l_i=2S, \qquad n_l=r/(2S).
  \label{upper_bound_rows_saturated}
\end{equation}
On the other hand, an upper bound for the second sum in Eq.~(\ref{casimir_young}) is found using the Cauchy-Schwartz inequality on the $n_c-$component vectors $(c_1,\ldots,c_n)$ and $(1,\ldots,1)$:
\begin{equation}
  (c_1^2+\ldots+c_n^2)(1+\ldots+1) \ge (c_1+\ldots+c_n)^2.
  \label{cauchy-schwarz_columns}
\end{equation}
The number of columns in the Young tableau $n_c$ is bounded by $n_c=\text{max}(\{l_i\})\le 2S$, and their sum is $\sum c_i = r$.
Hence, Eq.~(\ref{cauchy-schwarz_columns}) gives
\begin{equation}
  \sum_i^{n_c}c_i^2 \ge \frac{r^2}{n_c}\ge\frac{r^2}{2S}.
  \label{lower_bound_columns}
\end{equation}
This bound is saturated by
\begin{equation}
  c_i = r / (2S), \qquad n_c = 2S.
  \label{lower_bound_columns_saturated}
\end{equation}
Inserting Eqs.~(\ref{upper_bound_rows}) and (\ref{lower_bound_columns}) in Eq.~(\ref{casimir_young}) we get
\begin{equation}
  C \le \frac{N+2S}{2} \left(-\frac{r^2}{2SN}+r\right)\le \frac{NS(2S+N)}{4},
  \label{upper_bound_casimir}
\end{equation}
where the upper bound is obtained for $r=NS$. Together with Eqs.~(\ref{upper_bound_rows_saturated}) and (\ref{lower_bound_columns_saturated}), this precisely corresponds to the Young tableau of Fig.~\ref{young}.
Its Casimir eigenvalue is most easily computed using Weyl's formula [Eq.~(\ref{weyl})] and Eq.~(\ref{dynkin}), obtaining Eq.~(\ref{casimir_thisrep}), which saturates the upper bound of Eq.~(\ref{upper_bound_casimir}). Alternatively, the Casimir eigenvalue can be obtained from Eq.~(\ref{casimir_young}), and $r=NS$, $l_1=l_2=\ldots=l_{N/2}=2S$ and $c_1=c_2=\ldots=c_{2S}=N/2$.

Finally, we observe that, since the variables $\{l_i\}$ and $\{c_i\}$ in Eq.~(\ref{casimir_young}) are positive integers, as soon as we deviate from the solution maximizing $C$, we decrease the Casimir eigenvalue by a {\it finite} integer amount. In other words, there is a finite gap $O(1)$ in the eigenvalues of the quadratic Casimir operators between the subspace of the representation of Fig.~\ref{young} and the other irreducible representations arising from the tensor product of $2S$ self-adjoint antisymmetric representations.

\section{Systematic errors}
\label{app:dtau_errors}
\begin{figure}
  \centering
  \includegraphics[width=\linewidth]{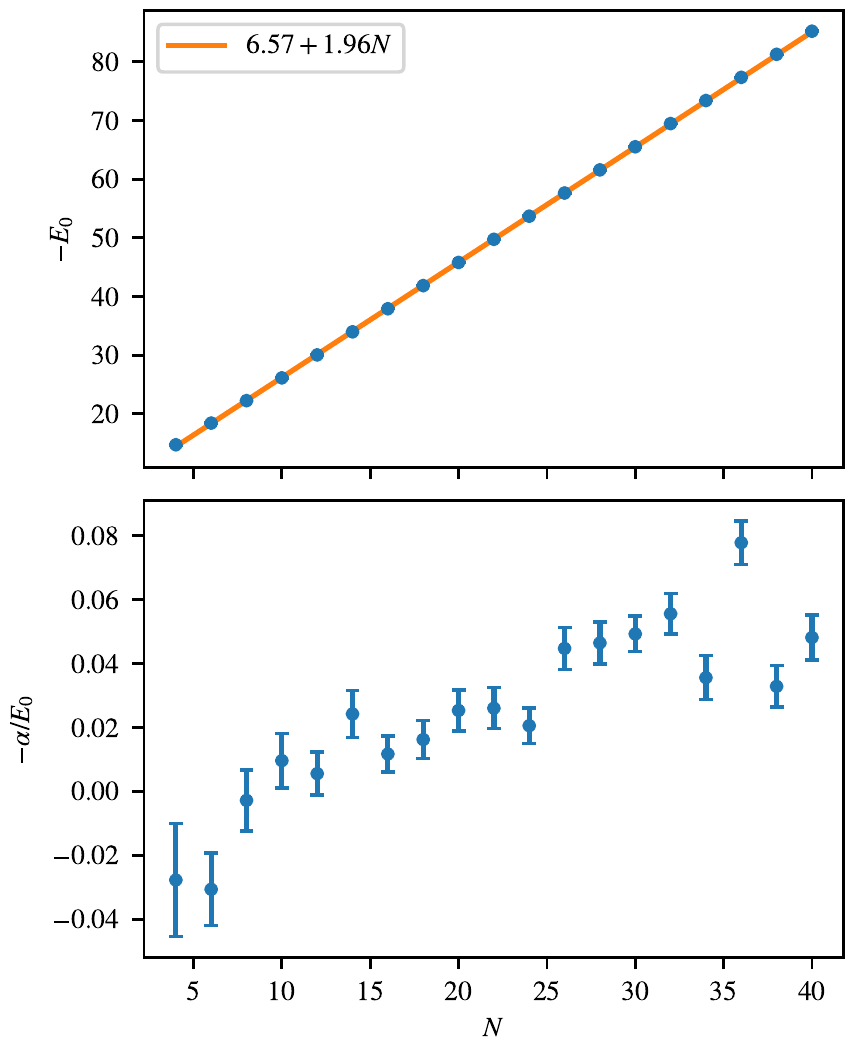}
  \caption{Scaling of systematic $\Delta_\tau$ error for $L=4$,  $\Theta = 2$, and  $S=1/2$. For each point, we simulated with a range of different values for $\Delta\tau$ and fitted the energy to $E(\Delta_\tau) = E_0 + \alpha {\Delta_\tau}^2$. }
  \label{fig:dtau_errors}
\end{figure} 

In this  appendix  we  show  that there is no  \textit{explicit}  dependence   on  the magnitude  of  the  Trotter error   as  a  function of  $N$.    
To keep    the notation simple, we  will show  this on the  basis  of  the  $S=1/2$  Hamiltonian  where  $\hat{H}_{\text{Casimir}}$   [see  Eq.~\ref{HQMC}]   as  well  as  the orbital index  can be  omitted: 
\begin{equation}
  \begin{split}
    \hat{H}_{\text{QMC}} =&\hat{H}_J + \hat{H}_U \\
    =
    &- \frac{J}{2N}  \sum_{\langle i,j \rangle } \left\{ \hat{D}_{i,j}, \hat{D}^{\dagger}_{i,,j}   \right\} 
      + \frac{U}{N} \sum_{i}\left(\nop{i}-\frac{N}{2}\right)^2 
  \end{split}
  \label{HQMC_1}
\end{equation}
In this  appendix,  we  have  normalized  the  Hamiltonian by  the  factor  $\frac{1}{N} $,    such  that   total  energy  differences   defining,  e.g.,  the spin gap  [$E_0(S=1)  - E_0$] remain  constant in  the  large-$N$  limit.     In particular,  with  the mean-field   ansatz    $\chi_{i,j}  =  \frac{1}{N} \langle  \hat{D}_{i,j} \rangle $   corresponding  to the  Affleck and  Marston  saddle  point \cite{Affleck88},  the  Hamiltonian  reads: 
\begin{equation}
	\hat{H}_{\text{MF}}   = - \frac{J}{2}  \sum_{\langle i,j \rangle }  \chi_{i,j} \hat{D}^{\dagger}_{i,,j}  + \overline{\chi_{i,j}} \hat{D}^{}_{i,,j}.
\end{equation} 
In this  large-$N$  limit,    one  will   check explicitly   that  the  spin gap  on  a  finite lattice is  $N$ independent, and   that  the   energy  is  extensive  in the   volume, $V$, and    in  $N$.  Since    gaps   are  $N$ independent,  at  least  in the large-$N$  limit,  it  makes  sense  comparing  results at  different $N$  but  at  constant  temperature  or   projection parameter. 

In the formulation of the AF QMC method, one introduces a checkerboard decomposition, where the interaction terms are grouped into disjoint families of commuting operators.
  This factorization introduces a Trotter discretization error, whose $N-$dependence we estimate as follows.
To  render  the calculation as  simple as possible,  we  will consider as an illustration a one-dimensional chain.  In this case,   the  
checkerboard  decomposition  in  even  and  odd  bonds,  $b$,     allows  us  to  write  the  Hamiltonian  as:
\begin{equation}
\hat{H}  =    \underbrace{\sum_{b \in A } \hat{h}_b}_{\equiv \hat{H}_A}    +   \underbrace{\sum_{b \in B } \hat{h}_b}_{\equiv \hat{H}_B}.
\end{equation}
Both   $ \hat{H}_A$ and  $\hat{H}_B $  are  sums  of  commuting  terms.    $\hat{h}_b $    corresponds  to a  local  Hamiltonian, 
such  that it  is  extensive  in $N$  but  intensive  in volume.   

An  explicit  form of   $\hat{h}_b $   on a  bond  with  legs  $i$,$j$   in the  fermion 
representation   would read:   $ \frac{1}{N} \left\{ \hat{D}_{(i,j)}, \hat{D}^{\dagger}_{(i,j)}  \right\}  $.   Note  that to keep   calculations as  simple  as possible   we,   implicitly  consider  a  one-dimensional lattice  in  which  $\hat{H}_A$  is  a  sum  of  commuting terms.  For the  two-dimensional case,  
the  checkerboard    bond  decomposition  necessitates  four terms.  
We  will  use  the   symmetric  Trotter  decomposition, 
\begin{equation}
	  e^{-\Delta  \tau \hat{H}  + \Delta \tau^3 \hat{R}_3}   =  e^{-\frac{\Delta  \tau}{2} \hat{H}_A}  e^{-\Delta  \tau \hat{H}_B} e^{-\frac{\Delta  \tau}{2}\hat{H}_A}  
	  +  {\cal O } \left(  \Delta \tau^5\right) 
\end{equation}
with  $\hat{R}_3  =  \left( \left[ \hat{H}_A, \left[ \hat{H}_A, \hat{H}_B \right] \right]  +  2 \left[ \hat{H}_B, \left[ \hat{H}_B, \hat{H}_A \right] \right] \right)/24 $. 
Since   $\hat{H}_A$  and  $\hat{H}_B $   are  sums  of  local  operators,      $\hat{R}_3$  is  also  a  sum of  local operators.  Hence,   $\hat{R}_3$  is  
extensive in the  volume.      By  explicitly  computing  the commutators,  one  will  also  show,    that  $\hat{R}_3 $  is  extensive  in  $N$.        Hence  
$\hat{R}_3 $    scales  as  $\hat{H} $.     Note  that for  non-local Hamiltonians   considered in Ref.~\cite{WangZ20}, this  does not apply.   We  can 
now  compute   the  corrections to  the  free  energy: 
\begin{equation}
	F_{\text{QMC}} =    F  -    \Delta\tau^{2} \frac{\text{Tr} e^{-\beta  \hat{H}} \hat{R}_3  } { \text{Tr} e^{-\beta  \hat{H}}  }  +  {\cal O} ( \Delta  \tau^4). 
 \end{equation}
Hence  the  quantity  plotted  in  Fig.~\ref{fig:dtau_errors}   corresponds  to  
\begin{equation}
	 \langle  \hat{R}_3  \rangle  /  \langle  \hat{H} \rangle. 
\end{equation}
It  is  intensive in   $N$ and $V$, such  that  it  has  a  well  defined  value  in the large-$N$ limit.  

\section{Bounds on the bond observable}
\label{app:bounds}
In this appendix, we discuss a lower and upper bound for a bond observable $\sum_a \hat{S}^{(a)}_i \hat{S}^{(a)}_j$, where $i$ and $j$ are two distinct lattice sites, not necessarily nearest neighbor.
The bond observable can be expressed as
\begin{multline}
  \sum_a\hat{S}^{(a)}_i \hat{S}^{(a)}_j =
  \frac{1}{2}\sum_a \left(\hat{S}^{(a)}_i + \hat{S}^{(a)}_j\right)\left(\hat{S}^{(a)}_i + \hat{S}^{(a)}_j\right)\\
  -\frac{1}{2}\sum_a\hat{S}^{(a)}_i\hat{S}^{(a)}_i
  -\frac{1}{2}\sum_a\hat{S}^{(a)}_j\hat{S}^{(a)}_j.
\label{SiSj_decomposition}
\end{multline}
With the choice of Eq.~(\ref{Ta_norm}),
the first term on the right-hand side of Eq.~(\ref{SiSj_decomposition}) is the quadratic Casimir element $\hat{C}_{2,\Gamma_i\otimes \Gamma_j}$ of the tensor product of the two \suna\ representations $\Gamma_i$ and $\Gamma_j$, at lattice sites $i$ and $j$ [compare with Eq.~(\ref{casimir_def})].
The spectrum of  $\hat{C}_{2,\Gamma_i\otimes \Gamma_j}$ consists in the eigenvalues of the quadratic Casimir operator of all irreducible representations to which $\Gamma_i\otimes\Gamma_j$ reduces.
An upper bound is readily found by the maximally symmetric composition of $\Gamma_i$ and $\Gamma_j$, which corresponds to a Young tableau with $N/2$ rows and $4S$ columns; the proof is identical to that of Appendix~\ref{app:casimir_max}.
Being a square of a hermitian operator, $\hat{C}_{2,\Gamma_i\otimes \Gamma_j} \ge 0$.
Such lower bound is saturated by the totally antisymmetric composition of $\Gamma_i$ and $\Gamma_j$, which corresponds to the trivial $S=0$ representation.
The second and third term on the right-hand side of Eq.~(\ref{SiSj_decomposition}) are the quadratic Casimir operators of the \suna\ representation considered here, and take the value given in Eq.~(\ref{casimir_thisrep}).
Inserting the bounds on $\hat{C}_{2,\Gamma_i\otimes \Gamma_j}$ discussed above in Eq.~(\ref{SiSj_decomposition}), we obtain
\begin{multline}
  -C(N,S) \le \Braket{\sum_a\hat{S}^{(a)}_i \hat{S}^{(a)}_j} \\ \le C(N, 2S)/2-C(N,S) = \frac{NS^2}{2}.
  \label{SiSrj_bounds}
\end{multline}


\begin{thebibliography}{56}
\bibitem [{not()}]{note-weyl}%
  \BibitemOpen
  \href@noop {} {}\bibinfo {note} {{The Weyl's formula for the Casimir element,
  Eq.~(\ref{weyl}), should not be confused with the Weyl's formula for the
  dimension of an irreducible representation,
  Eq.~(\ref{weyl_dimension}).}}\BibitemShut {Stop}%
\end{thebibliography}
\end{document}